\documentclass[]{aa}
\bibpunct{(}{)}{;}{a}{}{,} %% natbib format for A&A and ApJ
\usepackage{graphicx}
\usepackage{txfonts}
\usepackage{url}
\usepackage{amssymb,amsmath,amsfonts}
\usepackage[section]{placeins}
\usepackage{verbatim}
\usepackage{enumitem}
\usepackage{epsfig}
\usepackage{mathptmx}
\usepackage{booktabs,lscape,longtable}
\usepackage{comment}
\usepackage{color}
\usepackage[caption=false]{subfig}
\usepackage[draft]{hyperref}
\usepackage{natbib,twoopt}
\usepackage{bm}
\newcommand\blfootnote[1]{%
  \begingroup
  \renewcommand\thefootnote{}\footnote{#1}%
  \addtocounter{footnote}{-1}%
  \endgroup}

\begin{document} 
   \title{The Brown dwarf Atmosphere Monitoring (BAM) Project I:\\The largest near-IR monitoring survey of L \& T dwarfs}

   \titlerunning{The Brown dwarf Atmosphere Monitoring (BAM) Project I}
   \authorrunning{Wilson, Rajan \& Patience}

   \author{P. A. Wilson \inst{1}
          \and
          A. Rajan\inst{2}
          \and
          J. Patience\inst{1,2}
          }

   \institute{Astrophysics Group, School of Physics, University of Exeter, Stocker Road, Exeter EX4 4QL, UK\\
              \email{paw@astro.ex.ac.uk}
         \and
             School of Earth \& Space Exploration, Arizona State University, Tempe, AZ USA 8528\\
             }

   \date{Received November 6, 2013; accepted April 17, 2014}

% \abstract{}{}{}{}{} 
% 5 {} token are mandatory
 
  \abstract
   {Using the SofI instrument on the 3.5~m New Technology Telescope, we have conducted an extensive near-infrared monitoring survey of an unbiased sample of 69 brown dwarfs spanning the L0 to T8 spectral range, with at least one example of each spectral type. Each target was observed for a 2 -- 4 hour period in the $J_{\rm s}$-band, and the median photometric precision of the data is $\sim$~0.7\%. A total of 14 brown dwarfs were identified as variables with min-to-max amplitudes ranging from 1.7\% to 10.8\% over the observed duration. All variables satisfy a statistical significance threshold with a $p$-value $\leq 5$\% based on comparison with a median reference star light curve. Approximately half of the variables show pure sinusoidal amplitude variations similar to 2MASSJ2139+0220, and the remainder show multi-component variability in their light curves similar to SIMPJ0136+0933. It has been suggested that the L/T transition should be a region of a higher degree of variability if patchy clouds are present, and this survey was designed to test the patchy cloud model with photometric monitoring of both the L/T transition and non-transition brown dwarfs. The measured frequency of variables is $13^{+10}_{-4}$\% across the L7 -- T4 spectral range, indistinguishable from the frequency of variables of the earlier spectral types ($30^{+11}_{-8}$\%), the later spectral types ($13^{+10}_{-4}$\%), or the combination of all non-transition region brown dwarfs ($22^{+7}_{-5}$\%). The variables are not concentrated in the transition, in a specific colour, or in binary systems. Of the brown dwarfs previously monitored for variability, only $\sim60$\% maintained the state of variability (variable or constant), with the remaining switching states. The 14 variables include nine newly identified variables that will provide important systems for follow-up multi-wavelength monitoring to further investigate brown dwarf atmosphere physics.}

   \maketitle
%
%________________________________________________________________
\section{Introduction}
The L, T, and Y-type brown dwarfs represent a link between the coolest stars and giant planets. Many brown dwarfs are even cooler than currently observable exoplanetary atmospheres (e.g. HR8799b, HD189733b; \citealt{barman11}, \citealt{sing09,sing11}). The recently discovered Y dwarfs \citep{cushing11} approach the temperature of Jupiter. Since brown dwarfs never achieve a stable nuclear burning phase, they cool throughout their lifetimes, and temperature, rather than mass, is the dominant factor in defining the spectral sequence. As they cool, their atmospheres undergo changes in the chemistry and physical processes that sculpt their emergent spectra. While spectroscopy can be used to investigate atmospheric constituents and chemistry, photometric monitoring is an effective means to search for evidence of surface brightness inhomogeneities caused by cloud features, storms, or activity.\blfootnote{Based on observations made with ESO Telescopes at La Silla Observatory under programme ID 188.C-0493.}

The transition region from late-L to early-T encompasses a particularly interesting change in physical properties, as the atmospheres transform from dusty to clear over a narrow effective temperature range, and the observed infrared colours reverse from red to blue. This is predicted to be an effect of the formation and eventual dissipation of dusty clouds in brown dwarf atmospheres \citep{chabrier00,marley02,burrows06}. Broadly, as brown dwarfs cool through the spectral sequence, the lower temperatures allow more complex molecules to form, resulting in condensate clouds. When the temperature is cool enough, large condensate grains cannot remain suspended high in the atmosphere and sink below the observable photosphere, allowing methane and molecular hydrogen to become the dominant absorbers. Although there are several existing models for condensate cloud evolution, most cannot easily explain the rapid colour change from red to blue over the L-to-T transition. A systematic survey of variability in brown dwarfs including both L/T transition objects and comparison hotter/cooler objects is required to search for differences in the structure of condensate clouds in this important regime.

Existing photometric monitoring campaigns of brown dwarfs have been conducted at different wavelengths: optical bands (e.g. \citealt{tinney99} and \citealt{koen13}), near-IR bands (e.g. \citealt{artigau03}, \citealt{khandrika13} and \citealt{buenzli13}), mid-IR \citep[e.g.][]{morales-calderon06}, and radio frequencies \citep[e.g.][]{berger06}. From small ($<20$ objects) initial samples of ultracool field dwarfs, frequencies of variables ranged from 0\% to 100\% \citep[e.g. summary in][]{bailer-jones05}, and results from larger studies ($\sim25$ objects) have measured the frequency of variables to be in the range of 20\% to 30\% \citep[e.g.][]{khandrika13, buenzli13}. Examples of objects that vary in multiple wavebands have been identified (e.g. 2MASS J22282889-4310262 \citealt{clarke08, buenzli12}, SIMP J013656.5+093347.3 \citealt{artigau09}, 2MASS J21392676+0220226 \citealt{radigan12}), as well as objects recorded as variable in one wavelength range, but not another (e.g. 2MASS J15344984-2952274, \citealt{koen04b}). A small set of variable sources have been monitored contemporaneously at multiple wavelengths, with the combined results being used to infer the vertical extent of atmospheric features and to investigate atmospheric circulation patterns \citep[e.g.][]{buenzli12}. Given the unique probe of the atmospheric structure that multi-wavelength observations provide, it is essential to identify a larger set of known variables across a broad range of effective temperatures.

Most monitoring programs have involved observation sequences spanning a few hours, but some studies have searched for longer timescale variations \citep[e.g.][]{gelino02,enoch03}. A time scale of a few hours is well-matched to a search for rotation-modulated variability, since expected rotation periods are $\sim2$ -- 12 hours for L and T dwarfs, considering the range of measured $v\sin i$ values (10 -- 60~km/s for L dwarfs and 15 -- 40~km/s for T dwarfs -- \citealt{zapatero06}) and the $\sim$~0.08 -- 0.10~$M_{\odot}$ radius of these objects from evolutionary models at the age of the field \citep{baraffe03}. Periodogram analysis of some variables has shown clear peaks associated with periods in the range of $\sim2$ -- 8~hours \citep[e.g.][]{clarke08,radigan12} which is consistent with an atmospheric feature rotating into and out of view. Other variables exhibit multi-component light curves \citep[e.g.][]{artigau09} that are suggestive of a rapid evolution of atmospheric features.

To investigate the variability of brown dwarfs across the full L-T spectral sequence, we have performed a large-scale $J_{\rm s}$-band photometric monitoring campaign of 69 field brown dwarfs with the SofI instrument on the 3.5\,m New Technology Telescope (NTT). This survey is a part of the BAM (Brown dwarf Atmosphere Monitoring) project. In Section \ref{sample}, the properties of the sample, including magnitudes, spectral types, and companions are summarised. Details of the observations are reported in Section \ref{obs}, followed by the data reduction procedure, and methodology used to characterise each target as variable or constant in Section \ref{reduction}. Section~\ref{results} presents the results of the program and a comparison to previous variability studies. Finally, we discuss the sensitivity of the BAM survey and investigate possible correlations between variability and various observables such as spectral type, colour and binarity in Section~\ref{discussion}. The results are summarized in Section~\ref{conclusion}.

%__________________________________________________________________

\section{The BAM sample}
\label{sample}
\begin{figure}
\includegraphics[width=84mm,trim=0in 0in 0in 0]{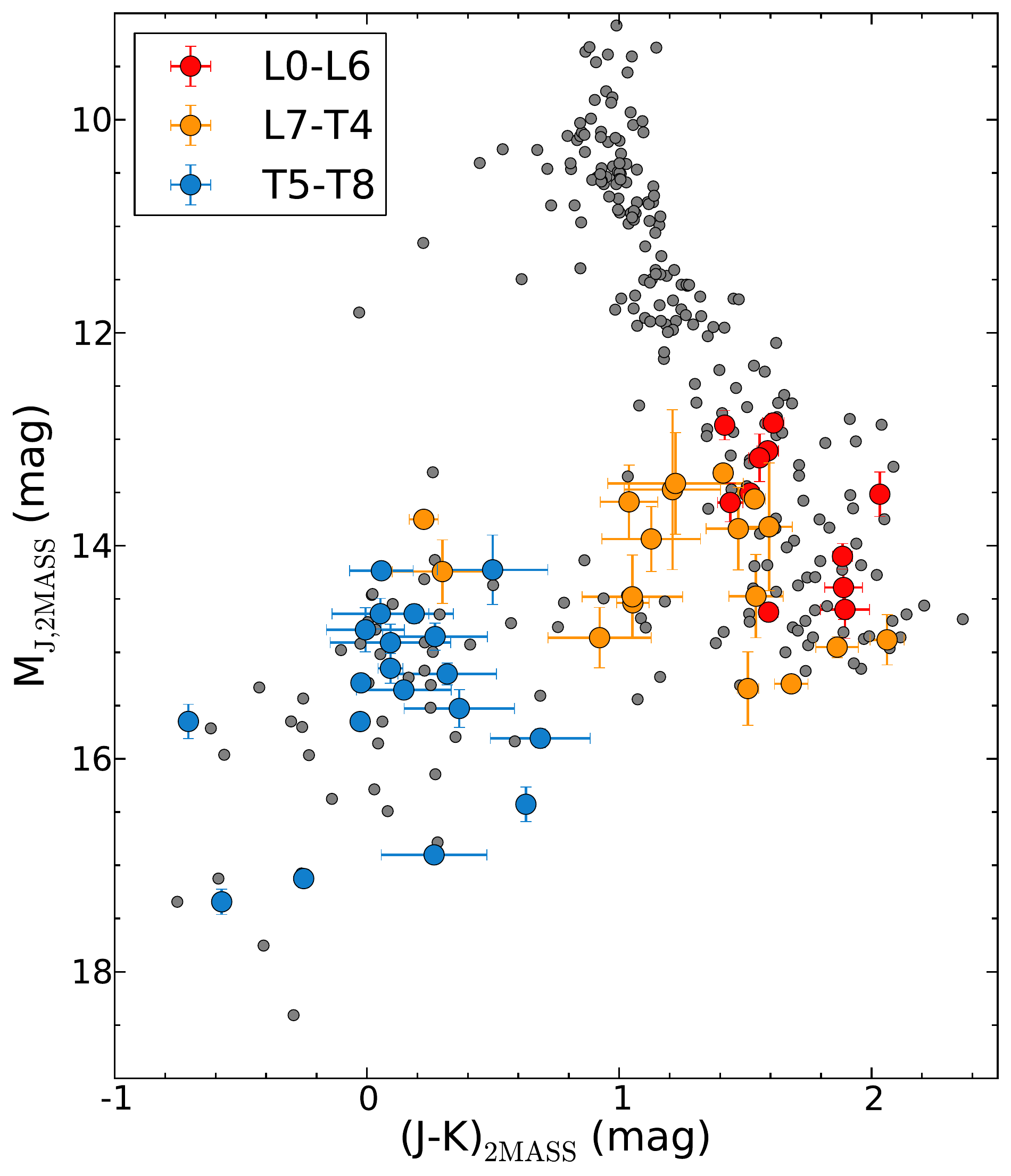}
 \caption{Colour-magnitude diagram of the M-L-T spectrum (small grey circles). All brown dwarfs with known parallax in the BAM sample are overplotted, with red representing the L dwarfs, yellow the L/T transition dwarfs, and blue the T dwarfs (see Table \ref{tabl:LdwarfSamp} and \ref{tabl:TdwarfSamp}). Half spectral types have been rounded down in the study. The photometry and parallaxes for the field M-L-T objects are from \citet{dupuy12}.}
  \label{cmd}
\end{figure}

\begin{figure*}%[!ht]
\centering
\includegraphics[width=85mm]{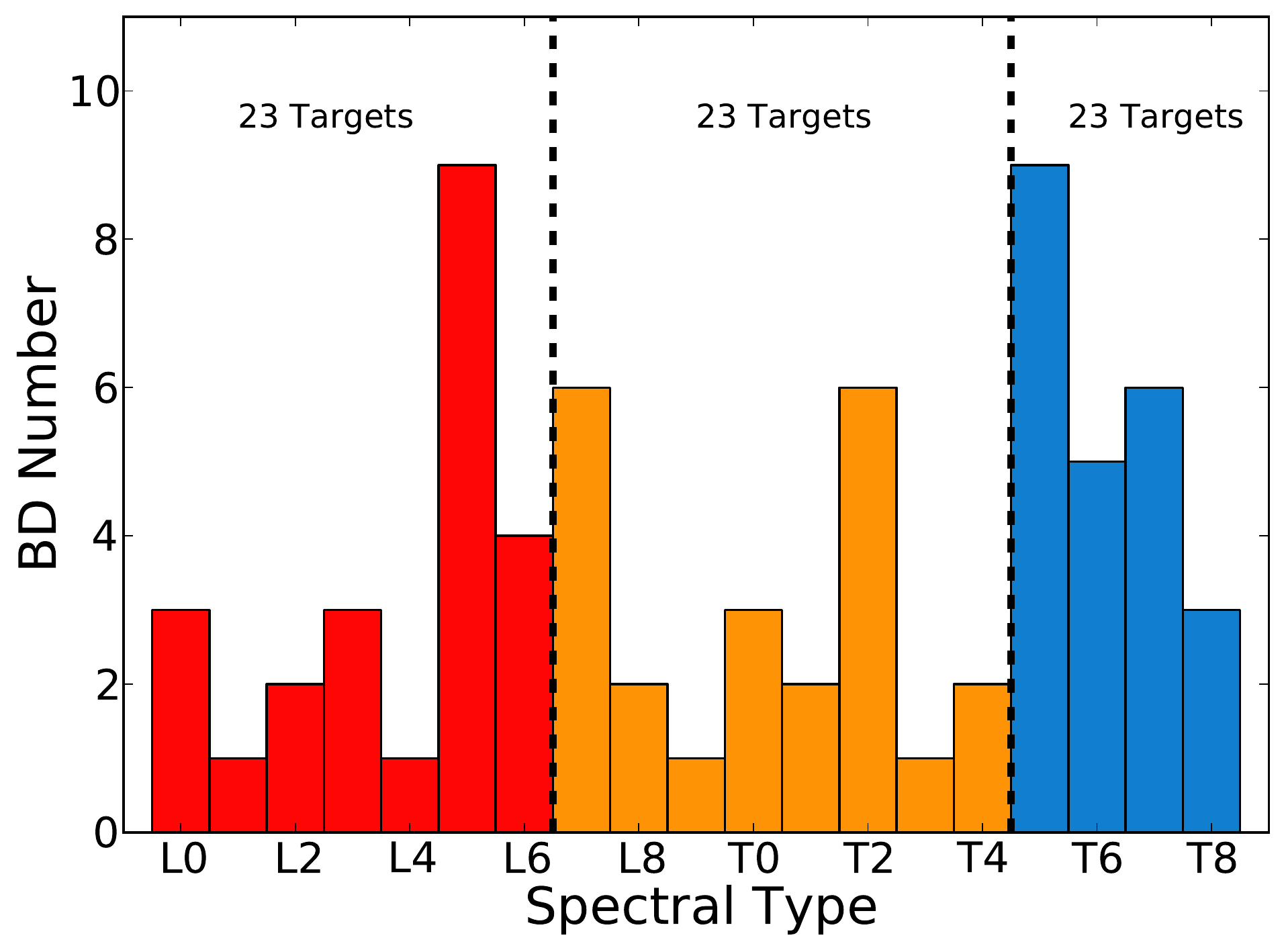}
\includegraphics[width=86mm]{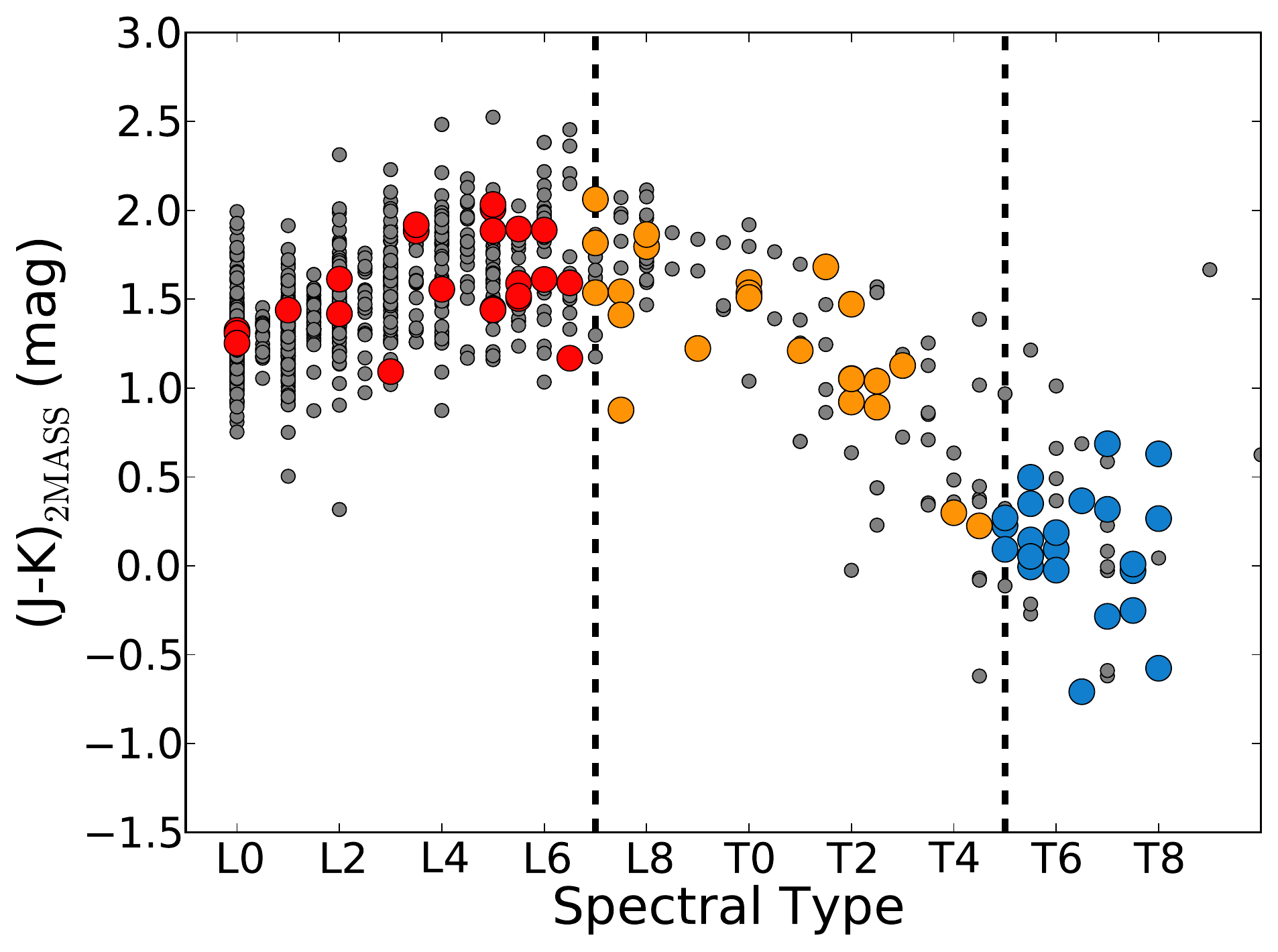}
 \caption{Diagram on the left shows a histogram of the sample across their respective spectral classes, whilst on the right is a colour-colour diagram showing the J-K colours of the same (coloured circles) overplotted on the full brown dwarf L-T spectral sequence (small grey circles). The L/T transition is indicated by the dashed lines defined in \citet{golimowski04}.}
  \label{spectral_type}
\end{figure*}

The 69 objects in the BAM sample were drawn from the brown dwarf archive (\url{dwarfarchives.org}) and were selected to span the full sequence of L- and T-spectral types from L0 to T8. An equal proportion of targets with spectral types above, across and below the L/T transition region were included. In this paper, we consider the L-T transition to range from L7~--~T4, following \cite{golimowski04}. Spectral types including a fractional subtype have been rounded down - for example, an L6.5 is considered L6 for the statistics. For the 48 targets with parallax measurements \citep[e.g.][]{dupuy12,faherty12}, a colour-magnitude diagram was constructed and is shown in Figure~\ref{cmd}. The histogram of target spectral types and a plot of the colour as a function of spectral type are shown in Figure~\ref{spectral_type}. The spectral types are based on IR spectroscopy for 54 targets and on optical spectroscopy for the remaining 15 targets that lacked an IR spectral classification. The spectral types, parallaxes, and apparent 2MASS magnitudes of the targets are listed in Table~\ref{tabl:LdwarfSamp} for L dwarfs and Table~\ref{tabl:TdwarfSamp} for T dwarfs. 

%__________________________________________________________________

%__________________________________________________________________

%__________________________________________________________________

Additional factors that influenced the target selection were the magnitudes and coordinates. To obtain high signal-to-noise individual measurements, the targets were limited to objects with magnitudes brighter than $J\sim$16.5~mag. To avoid observations at high airmass, the target declinations were limited to South of +20~degrees. Pairs of targets were observed for sequences of 2 to 4 hours, which also impacted the range of target coordinates observed each observing run. 

The majority of the sample, 47 targets, have been observed in programs designed to detect binary companions with radial velocity variations \citep{blake10}, or spectra showing features of different spectral types \citep{burgasser10}, or high angular resolution imaging \cite[i.e.][]{buoy03,mccaughrean04,burgasser03a,burgasser05,burgasser06,R06,reid08,looper08}. Based on the companion search programs reported in the literature, a total of 12 targets are members of spatially resolved binary pairs, and the results of all the binary searches are reported in Table \ref{tabl:LdwarfSamp} and \ref{tabl:TdwarfSamp}. The proportion of binaries in this sample is comparable to the overall brown dwarf binary frequency \citep{burgasser03a}, indicating that the sample is not biased in the level of multiple systems included. All of the binaries in the sample have separations less than the seeing limit, so the photometric measurements in this study record the combined flux from both components. A variable brown dwarf that is part of an unresolved binary system can be more difficult to detect, as the variability is diluted by the non variable companion.

Previous observations designed to search for photometric variability have been reported for approximately half the sample -- 34 targets -- and cover optical \citep{gelino02,koen13}, near-IR \citep{enoch03,koen04b,koen05b,clarke08,khandrika13,buenzli13}, and radio \citep{berger06} wavelengths. It is important to note that the different variability monitoring studies apply different criteria to categorise a target as variable or constant, and a range of observation wavelengths have been employed. Most of the previous monitoring has been conducted over timescales of hours similar to this program, though a few studies covered longer timescales with lower cadence measurements (e.g. \citealt{gelino02}, \citealt{enoch03}). 

\section{Observations}
\label{obs}
The observations took place from 4 - 11 October 2011 and 3 - 9 April 2012 with the SofI (Son of ISAAC) instrument \citep{moorwood98} mounted on the NTT (New Technology Telescope) at the ESO La Silla observatory. Observations were performed in the large field imaging mode that has a pixel scale of 0$\farcs288$~px$^{-1}$ and a field-of-view of $4\farcm92 \times 4\farcm92$. During the first observing run, some of the targets were observed in both the $J_{\rm s}$-band and $K_{\rm s}$-band, but only the $J_{\rm s}$-band was used during the second run. As a consequence, six of the targets from the first run have $J_{\rm s}$-band data with lower cadence. The $J_{\rm s}$ filter (1.16-1.32~$\mu m$) was used to avoid contamination by the water band centred at $1.4~\mu m$ that would have otherwise affected the photometry. An increase in the telluric water column would have caused an anti-correlation between the brightness of the brown dwarfs and the reference stars in the $J$-band, since an increase in the water column will decrease the flux from the reference stars to a greater extent compared to the brown dwarfs that have deep intrinsic water bands. The $J_{\rm s}$ data should not suffer from this effect.

Three sets of two target fields were observed most nights, alternating between each target roughly every 15~min over a $\sim3.5$~hour window. This procedure allowed six targets to be observed every night. During clear conditions, the observations had a detector integration time (DIT) of 5s, with three DITs (NDIT) taken and averaged together with about 25 exposures in each observing block. During poorer conditions, such as the presence of cirrus clouds, and for fainter objects, the exposure times were increased. The flux was kept below 10,000 ADUs for the brightest targets in the field to prevent any non-linearity effects. 

\section{Data Reduction and Photometry}
\label{reduction}

%__________________________________________________________________

\begin{table*}
\tiny{\centering
\caption{L dwarf Sample.}
\label{tabl:LdwarfSamp}
\begin{tabular}{ccccccc}
\hline
Target Name    & Spectral    &	Parallax		&	$J_{2MASS}$ 	& Binary/  & Instrument &	References \\
			& Type	&	(mas)		&   (mag)  & Single 	&		     & \\
\hline
\hline
2MASS J00165953-4056541 	& L3.5* & 				    & 15.316 $\pm$ 0.061 & 	S	&  SpeX spectra &	K08, BC10 \\
2MASS J00184613-6356122  	& L2* 	&  				    & 15.224 $\pm$ 0.044 & 		& 			 & R08\\
2MASS J01062285-5933185  	& L0* 	&   				& 14.33 $\pm$ 0.035 & 		& 			& R08\\
2MASS J01282664-5545343     & L1    & 15.24 $\pm$ 7.9   & 13.775 $\pm$ 0.027 &      &       & K07, D07 \\
DENIS J0205.4-1159  	& L5.5 	& 50.6 $\pm$ 1.5	    & 14.587 $\pm$ 0.03  & B (409 mas) & HST imaging  & R06, K04, D02, B03\\
2MASS J02284355-6325052  	& L0 	&   				& 13.556 $\pm$ 0.028 & 		& 			& K07\\
2MASS J02572581-3105523 	& L8* 	&   				& 14.672 $\pm$ 0.039 & 		& 			& K08, R08 \\
2MASS J03185403-3421292 	& L7* 	&   72.9 $\pm$ 7.7	& 15.569 $\pm$ 0.055 & S & HST imaging & K08, F12, R08 \\
   						    & 		&				&				   & S & SpeX spectra & BC10 \\
2MASS J03400942-6724051  	& L7* 	&   				& 14.742 $\pm$ 0.032 & 		& 			& C07, K08\\
2MASS J03582255-4116060 	& L5* 	&   				& 15.846 $\pm$ 0.087 & 		& 			& R08 \\
2MASS J04070752+1546457 	& L3.5 	&   				& 15.478 $\pm$ 0.058 & 		& 			 & R08, B06 \\
2MASS J04390101-2353083 	& L6.5* &  110.4 $\pm$ 4.0 	& 14.408 $\pm$ 0.029 & S & HST imaging & C03, F12, R06 \\
2MASS J04455387-3048204 	& L2* 	&   78.5 $\pm$ 4.9	& 13.393 $\pm$ 0.026 & S & HST imaging & C03, F12, R06 	\\
2MASS J05233822-1403022  	& L5	&   				& 13.084 $\pm$ 0.024 & S & radial velocity & C03, W03, BCW10 \\
2MASS J06244595-4521548  	& L5*	&  	83.9 $\pm$ 4.5	& 14.48 $\pm$ 0.029   & S & HST imaging  & R08, F12, R06 \\
2MASS J08354256-0819237 	& L5* 	& 117.3 $\pm$ 11.2 	& 13.169 $\pm$ 0.024 & S & HST imaging & C03, And11, R06\\
   						    & 		&				    &				   & S & radial velocity & BCW10 \\
   						    & 		&				    &				   & S & SpeX spectra & BC10 \\
2MASS J09153413+0422045  	& L7*	&                   & 14.548 $\pm$ 0.03  & B (730 mas) & HST imaging & R08, R06 \\
2MASS J09310955+0327331	    & L7.5	&   				& 16.615 $\pm$ 0.138 & 		& 			& K04\\
2MASS J10043929-3335189  	& L4* 	& 54.8 $\pm$ 5.6	& 14.48 $\pm$ 0.035   & 		& 			& Giz02, And11\\
2MASS J10101480-0406499 	& L6* 	& 59.8 $\pm$ 8.1  	& 15.508 $\pm$ 0.059 & 		& 			& C03, F12 \\
2MASS J11263991-5003550 	& L6.5 	&   				& 13.997 $\pm$ 0.032 & 		& 			& F07, B06 \\
2MASS J11555389+0559577 	& L7.5 	& 57.9 $\pm$ 10.2  	& 15.66 $\pm$ 0.077   & S & SpeX spectra & K04, F12, BC10 \\
2MASS J12281523-1547342  	& L6 	& 44.8 $\pm$ 1.8	& 14.378 $\pm$ 0.03   & B (264 mas) & HST imaging & D97, K04, D12, B03\\
2MASS J13004255+1912354 	& L3 	&   				& 12.717 $\pm$ 0.022 & S & HST imaging & Giz00, B06, R08 \\
   						    &   	&				&				    & S & radial velocity & BCW10 \\
2MASS J13262981-0038314 	& L5.5 	& 50 $\pm$ 6		& 16.103 $\pm$ 0.071 & S & SpeX spectra & F00, K04, V04, BC10 \\
2MASS J15074769-1627386 	& L5.5 	& 136.4 $\pm$ 0.6	& 12.83 $\pm$ 0.027   & S & HST imaging & R00, K04, D02, R06 \\
2MASS J16322911+1904407 	& L8 	& 65.6 $\pm$ 2.1 	& 15.867 $\pm$ 0.07 & S & HST imaging &  K99, B06, D02, B03 \\
2MASS J19360187-5502322 	& L5* 	& 66.3 $\pm$ 5.4	& 14.486 $\pm$ 0.039 & S & HST imaging & R08, F12, R06 \\
SDSS J204317.69-155103.4 	& L9 	& 22.8 $\pm$ 4.7    & 16.625 $\pm$ 0.162 & S & SpeX spectra & C06, S13, BC10 \\
2MASS J22521073-1730134 	& L7.5 	& 63.2 $\pm$ 1.6	& 14.313 $\pm$ 0.029 & B (130 mas) & HST imaging & Ken04, D12, RL06 \\
2MASS J22551861-5713056 	& L5.5 	&                   & 14.083 $\pm$ 0.03   & B (120 mas) & HST imaging  & K07, R08 \\
2MASS J23224684-3133231     & L0*   & 58.6 $\pm$ 5.6    & 13.577 $\pm$ 0.027    &           &               & R08, F12 \\
\hline
Note: *Spectral classification using optical data\\
\end{tabular}
References: \citet{andrei11}[And11], \citet{buoy03} [B03], \citet{blake10} [BCW10], \citet{berger06} [B06], \citet{burgasser10} [BC10], \citet{cruz03} [C03], \citet{chiu06} [C06], \citet{cruz07} [C07], \citet{dahn02} [D02], \citet{deacon07} [D07], \citet{delfosse97} [D97], \citet{dupuy12}[D12], \citet{folkes07}[F07], \citet{faherty12} [F12], \citet{fan00} [F00], \citet{gizis00} [Giz00], \citet{gizis02} [Giz02], \citet{kendall04} [Ken04], \citet{kendall07} [K07], \citet{kirkpatrick99}[K99], \citet{kirkpatrick08}[K08], \citet{knapp04} [K04], \citet{reid00} [R00], \citet{R06} [R06], \citet{RL06} [RL06], \citet{reid08} [R08], \citet{smart13}[S13], \citet{vrba04} [V04].\\
}
\end{table*}

%__________________________________________________________________
\subsection{Processing the images}
For each image, basic data reduction steps consisting of correcting for the dark current and division by a flat field and sky subtraction were applied. Developing flat field images for the NTT/SofI instrument involved generating two different flats, a special dome flat and an illumination correction flat as documented by the observatory. The dome flat requires observations of an evenly illuminated screen with the dome lamp turned on and off in a particular set sequence. To correct for low frequency sensitivity variations across the array that are not completely removed by the dome flat, an illumination correction was applied. By observing the flux from a standard star in a grid pattern across the array, a low order polynomial was fitted to the flux measurements, allowing large scale variations across the array to be characterised and removed. Flat field images were produced using the IRAF\footnote{IRAF is distributed by the National Optical Astronomy Observatories, which are operated by the Association of Universities for Research in Astronomy, Inc., under cooperative agreement with the National Science Foundation.} scripts provided by the observatory\footnote{\url{http://www.eso.org/sci/facilities/lasilla/instruments/sofi/tools/reduction/sofi_scripts.html}}. As the flat fields are documented to be extremely stable over several months, a single set of flat fields were used for all the targets in a given run. 

For the SofI instrument, the dark frames are a poor estimate of the underlying bias pattern, which varies as a function of the incident flux. Consequently, the dark and bias are subtracted from the science frames through the computation of a sky frame, which also removes the sky background from the science data. Sky frames were generated by median combining the dithered science frames. The final calibration step involved measuring the offsets between the individual images and aligning all the science frames. The aligned frames within each $\sim15$~min interval were subsequently median combined. We compared the photometric uncertainties on the median combined images calculated using IRAF, to the standard deviation of the unbinned images within each bin. For most objects, the two methods for calculating uncertainties gave very similar results. The IRAF uncertainties on the median combined images were used for all the objects for consistency. Median combining the images before performing photometry rather than measuring the individual frames had the advantage of improving the centring, measurements of the full width at half maximum (FWHM), and the photometry of the fainter comparison stars in the field.

%__________________________________________________________________

\begin{table*}
\tiny{\centering
\caption{T dwarf Sample.}
\label{tabl:TdwarfSamp}
\begin{tabular}{ccccccc}
\hline
Target Name	& Spectral	&	Parallax		&	$J_{2MASS}$ 	& Binary/  & Instrument &	References \\
			& Type	&	(mas)		&   (mag)  & Single 	&		     & \\
\hline
\hline
2MASS J00345157+0523050 	& T6.5 	& 105.4 $\pm$ 7.5  	& 15.535 $\pm$ 0.045 & 		& 			 & B04, B06, F12 \\
2MASS J00501994-3322402 	& T7 	& 94.6 $\pm$ 2.4	& 15.928 $\pm$ 0.07   & 		& 			 & T05, B06, D12 \\
SIMP J013656.5+093347.3 	& T2.5 	&   				& 13.455 $\pm$ 0.03   & 		& 			 & A06 \\
2MASS J03480772-6022270 	& T7 	&   				& 15.318 $\pm$ 0.05   & S 	& HST imaging & B03, BK06 \\
2MASS J04151954-0935066 	& T8  	& 175.2 $\pm$ 1.7	& 15.695 $\pm$ 0.058 & S 	& HST imaging  & B02, B06, V04, BK06 \\
SDSS J042348.56-041403.5 	& T0  	& 72.1 $\pm$ 1.1	& 14.465 $\pm$ 0.027 & B (160 mas) & HST imaging & G02, B06, V04, B05 \\
2MASS J05103520-4208140 	& T5  	&   				& 16.222 $\pm$ 0.087 & 		& 			& L07 \\
2MASS J05160945-0445499 	& T5.5 	& 44.5 $\pm$ 6.5  	& 15.984 $\pm$ 0.079 & S 	& HST imaging & B03, F12, B06, BK06 \\
2MASS J05591914-1404488 	& T4.5 	& 96.6 $\pm$ 1	 	& 13.802 $\pm$ 0.024 & S 	& HST imaging & B00, B06, D02, BK03 \\
2MASS J07290002-3954043	    & T8 	& 126.3 $\pm$ 8.3  	& 15.92 $\pm$ 0.077   & 		& 			& L07, F12 \\
DENIS J081730.0-615520 	    & T6 	& 203 $\pm$ 13	    & 13.613 $\pm$ 0.024 & 		& 			& A10 \\
2MASS J09393548-2448279 	& T8 	& 187.3 $\pm$ 4.6 	& 15.98 $\pm$ 0.106   & 		& 			& T05, B06, B08 \\
2MASS J09490860-1545485 	& T2 	& 55.3 $\pm$ 6.6	& 16.149 $\pm$ 0.117 & ?B (weak candidate) & SpeX spectra & T05, F12, B06, BC10 \\
2MASS J10073369-4555147 	& T5 	& 71.0 $\pm$ 5.2  	& 15.653 $\pm$ 0.068  &  		& 			& L07, F12 \\
2MASS J10210969-0304197 	& T3 	& 29.9 $\pm$ 1.3 	& 16.253 $\pm$ 0.091  & B (172 mas) & HST imaging & L00, B06, T03, BK06 \\
2MASS J11145133-2618235 	& T7.5 	& 179.2 $\pm$ 1.4	& 15.858 $\pm$ 0.083  &  		& 			& T05, B06, D12 \\
2MASS J12074717+0244249	    & T0 	& 44.5 $\pm$ 12.2  	& 15.58 $\pm$ 0.071   & ?B (weak candidate) & SpeX spectra & H02, F12, B06, BC10 \\
2MASS J12171110-0311131 	& T7.5 	& 90.8 $\pm$ 2.2	& 15.86 $\pm$ 0.061   & S 	& HST imaging & B99, B06, T03, BK06 \\
2MASS J12255432-2739466 	& T6 	& 75.1 $\pm$ 2.5	& 15.26 $\pm$ 0.047   & B (282 mas) & HST imaging  & B99, B06, T03, BK03 \\
2MASS J12314753+0847331     & T5.5 	&   				& 15.57 $\pm$ 0.072  &   		& 			& B04, B06 \\
2MASS J12545393-0122474 	& T2 	& 84.9 $\pm$ 1.9	& 14.891 $\pm$ 0.035 & S & HST imaging & L00, B06, D02, BK06 \\
2MASS J14044941-3159329 	& T2.5 	& 42.1 $\pm$ 1.1	& 15.577 $\pm$ 0.062 & B (130 mas) & AO imaging &  L07, D12, L08 \\
2MASS J1511145+060742 	    & T2 	& 36.7 $\pm$ 6.4  	& 16.016 $\pm$ 0.079 & ?B (strong candidate) & SpeX spectra & A11, F12, BC10 \\
2MASS J15210327+0131426     & T2 	& 41.3 $\pm$ 7.2  	& 16.399 $\pm$ 0.102 & S & SpeX spectra & K04, F12, B06, BC10 \\
2MASS J15344984-2952274     & T5.5 	& 62.4 $\pm$ 1.3	& 14.9 $\pm$ 0.054     & B (65 mas) & HST imaging & B02, B06, T03, BK03 \\
2MASS J15462718-3325111 	& T5.5 	& 88 $\pm$ 1.9		& 15.631 $\pm$ 0.051 & S & HST imaging & B02, B06, T03, BK03 \\
2MASS J15530228+1532369 	& T7 	& 75.1 $\pm$ 0.9	& 15.825 $\pm$ 0.071 & B (349 mas) & HST imaging & B02, B06, D12, BK06  \\
2MASS J16241436+0029158 	& T6 	& 90.9 $\pm$ 1.2	& 15.494 $\pm$ 0.054 & S & HST imaging & S99, B06, T03, BK06 \\
2MASS J18283572-4849046 	& T5.5 	& 83.7 $\pm$ 7.7  	& 15.175 $\pm$ 0.056  & 		& 			& B04, F12, B06 \\
SDSS J204749.61-071818.3 	& T0 	& 49.9 $\pm$ 7.9  	& 16.85 $\pm$ 0.04* & S & SpeX spectra & K04, F12, B06, BC10 \\
2MASS J20523515-1609308 	& T1 	& 33.9 $\pm$ 0.8	& 16.334 $\pm$ 0.118  & ?B (weak candidate) & SpeX spectra & C06, D12, BC10 \\
                            &       &                   &                     & B (100.9 mas) & AO imaging &   S11 \\
2MASS J21392676+0220226 	& T1.5 	& 101.5 $\pm$ 2.0	& 15.264 $\pm$ 0.049 & ?B (strong candidate) & SpeX spectra &  R08, B06, S13, BC10 \\
2MASS J21513839-4853542 	& T4 	& 50.4 $\pm$ 6.7  	& 15.73 $\pm$ 0.074 &   		& 			& E05, F12, B06 \\
2MASS J22282889-4310262 	& T6.5 	& 94.0 $\pm$ 7.0  	& 15.662 $\pm$ 0.073 & S & HST imaging & B03, F12, B06, BK06 \\ 
ULAS J232123.79+135454.9 	& T7.5 	&   				& 17.04 $\pm$ 0.04* &   		& 			& S10, B10 \\
2MASS J23312378-4718274 	& T5 	&   				& 15.659 $\pm$ 0.068 &   		& 			& B04, B06 \\
2MASS J23565477-1553111 	& T5.5 	& 57.9 $\pm$ 3.5 	& 15.824 $\pm$ 0.057 & S & HST imaging & B02, B06, S13, BK03 \\
\hline
\end{tabular}
\\Note: *MKO to 2MASS conversion using transformations from \citet{stephens04}.\\
References: \citet{albert11} [A11], \citet{artigau06} [A06], \citet{artigau10} [A10], \citet{berger06} [B06], \citet{buoy03} [B03], \citet{burgasser99} [B99], \citet{burgasser00} [B00], \citet{burgasser02} [B02], \citet{burgasser03a} [BK03], \citet{burgasser04} [B04], \citet{burgasser05} [B05], \citet{burgasser06} [BK06], \citet{burgasser08} [B08], \citet{burgasser10} [BC10], \citet{burningham10} [B10], \citet{chiu06} [C06], \citet{dahn02} [D02], \citet{dupuy12}[D12], \citet{ellis05} [E05], \citet{faherty12} [F12], \citet{gelino02} [G02], \citet{hawley02} [H02], \citet{knapp04} [K04], \citet{leggett00} [L00], \citet{looper07} [L07], \citet{looper08} [L08], \citet{reid08} [R08], \citet{scholz03} [S03], \citet{scholz03} [S10], \citet{stumpf11} [S11], \citet{smart13} [S13], \citet{strauss99} [S99], \citet{tinney03} [T03], \citet{tinney05} [T05], \citet{vrba04} [V04].\\
}
\end{table*}
%__________________________________________________________________

\subsection{Generating the light curves}
Aperture photometry was carried out using the APPHOT package in IRAF. A median value of the FWHM was measured per image using all the stars in the field-of-view. A range of aperture radii were explored and the size of $1.5\times$FWHM was selected, as it minimised the root-mean-square (RMS) scatter of the reference star light curves that were created by dividing each reference star by the weighted mean of the remaining reference star light curves. The aperture was kept constant for all the stars in a single image, but was allowed to vary between individual images to account for variations in seeing. The variable aperture also yielded higher signal-to-noise measurements compared to a constant aperture, which would otherwise cause a loss in the flux measured within the aperture during poorer seeing conditions. We checked each target field to ensure that the photometry was not impacted by nearby astrophysical sources.

\begin{figure}
\centering
\includegraphics[width=84mm]{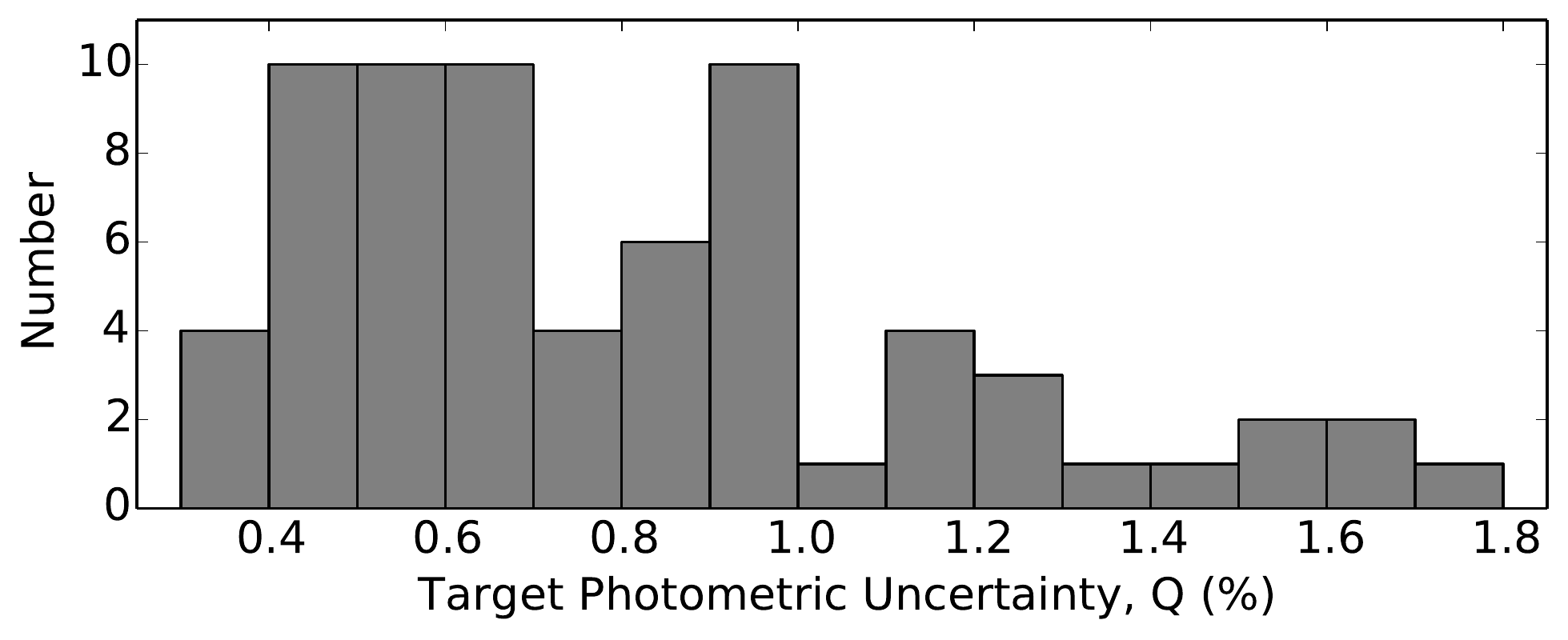}
 \caption{Target photometric uncertainty of the survey defined as the median value of the final target light curve uncertainties. The median target photometric uncertainty is $0.7\%$.} 
\label{phot_qual}
\end{figure}

The steps taken to generate the target light curves in the survey are given in the following list: 
\begin{itemize}
\item[$\bullet$] For each target, a list of reference star candidates was generated by considering all stars visible in the field of view, discarding stars with peak counts less than $20$~ADUs or greater than $10,000$~ADUs. These limits were imposed to ensure enough signal was present to accurately centre the aperture around the object and to ensure that none of the reference stars were in the non-linear regime of the detector. 
\item[$\bullet$] Reference candidates were trimmed by selecting up to 15 of the reference stars with the most similar brightness to the target.
\item[$\bullet$] Candidate reference star light curves were calculated by dividing each reference star by a weighted mean of the remaining reference stars.
\item[$\bullet$] Candidate reference stars with light curves exhibiting a standard deviation greater or equal to the median standard deviation for all reference star candidates were removed. 
\item[$\bullet$] A master reference light curve was subsequently created by median combining the normalised light curves of all the qualifying reference stars.
\item[$\bullet$] The final target light curve was produced by dividing the target brown dwarf flux by the weighted mean of all the qualifying reference stars. The light curve was normalised by dividing the light curve by the median flux value of the light curve.
\item[$\bullet$] The target and reference star light curves were all airmass de-trended by dividing the light curves by a second order polynomial fit to the relative flux of the master reference as a function of airmass.

\end{itemize}
\noindent The number of reference stars used for each target is given in Table~\ref{tabl:variables} and Table~\ref{tabl:constants}, with six to eight references being typical. The automatic selection process was applied uniformly throughout the entire sample of objects. The uncertainties were calculated using IRAF. The target photometric uncertainty ($Q$) is defined as the median value of the target light curve uncertainties. A histogram of the $Q$ values for each object is shown in Figure~\ref{phot_qual}, and the value for all targets are listed in Table~\ref{tabl:variables} and Table~\ref{tabl:constants}. The median $Q$ value for the entire survey is $0.7$\%.

%__________________________________________________________________
%__________________________________________________________________
%__________________________________________________________________
\begin{figure}
\centering
\includegraphics[width=84mm]{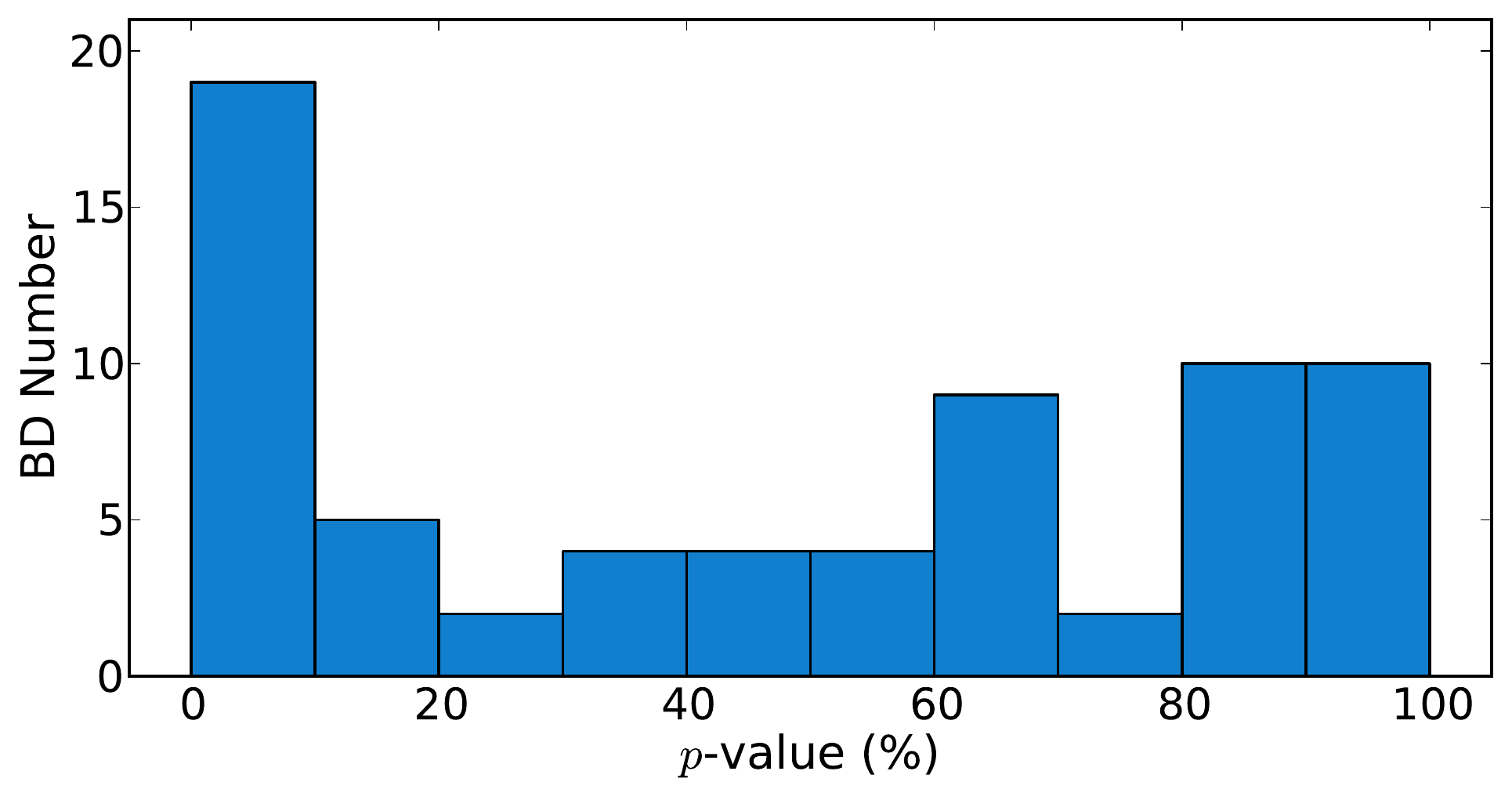}
 \caption{$p$-value histogram of the full brown dwarf sample. The objects in the first bin includes 16 targets with $p$-value~$\leq~5\%$, and three targets with $p$-value between 5 -- 10\%. Of the 16 targets with $p$-value~$\leq~5\%$ listed in Table~\ref{tabl:variables}, two targets are not listed since they failed the robust criterion ($\tilde{\eta}\ge1$). The large number of objects in the last two bins (80 to 100) is suggestive of conservative errorbars.} 
\label{pval}
\end{figure}

%__________________________________________________________________
%__________________________________________________________________
\begin{table*}
\tiny{\centering
\caption{Variables identified in this study.}
\label{tabl:variables}
\begin{tabular}{lccccccccc}
\hline
Object &   Spectral Type & Obs. Dur. (hours) & Refs. & DOF &  $\chi^2_\nu$ & $\tilde{\eta}$ & $Q$ (\%) & $p$-value (\%) &   Amplitude$^*$ (\%)\\
\hline
\hline
  \multicolumn{10}{c}{Variables with $p$-value~$\leq~5\%$ and $\tilde{\eta}\ge1.0$}\\ 
  \hline
  2MASS J01062285-5933185 & L0   & 3.14 & 7 & 5     & 4.4       & 1.3   & 1.07   & 0.12 & $4.3\pm1.2$\\
  2MASS J13004255+1912354 & L3    & 3.06 & 5 & 9    & 3.6       & 2.6   & 1.21    & 0.0 & $9.6\pm0.9$\\
  2MASS J03582255-4116060 & L5    & 1.97 & 6 & 4     & 4.2       & 1.9   & 0.88    & 0.2 & $4.8\pm1.2$\\
  2MASS J08354256-0819237 & L5    & 3.32 & 6 & 10    & 5.8       & 2.0   & 0.33    & 0.0 & $1.7\pm0.5$\\
  2MASS J22551861-5713056 & L5.5  & 3.16 & 3 & 6    & 4.0       & 2.5   & 1.55    & 0.0 & $9.4\pm1.6$\\
  2MASS J10101480-0406499 & L6    & 3.08  & 4 & 8    & 4.3       & 2.0   & 1.18    & 0.0 & $5.1\pm1.1$\\
  2MASS J04390101-2353083 & L6.5 & 2.53 & 8 & 5     & 2.7       & 1.7   & 0.50    & 1.8 & $2.6\pm0.5$\\
  2MASS J11263991-5003550 & L6.5  & 3.23 & 8 & 12    & 3.3       & 1.4   & 0.52    & 0.0 & $3.2\pm0.7$\\
  2MASS J12074717+0244249 & T0    & 2.76 & 7 & 7     & 2.1       & 1.4   & 0.91    & 4.2 & $5.2\pm1.1$\\
  2MASS J21392676+0220226 & T1.5  & 2.62 & 7 & 7     & 18.1      & 3.9   & 0.42    & 0.0 & $4.7\pm0.5$\\
  SIMP J013656.5+093347.3 & T2.5  & 2.94 & 5 & 7     & 2.3       & 2.2   & 0.72    & 2.3 & $3.0\pm0.6$\\
  2MASS J22282889-4310262 & T6.5  & 3.08 & 8 & 7     & 2.6       & 1.3   & 0.51    & 1.8 & $3.9\pm0.7$\\
  2MASS J00501994-3322402 & T7.0  & 2.63 & 6 & 6     & 5.6       & 3.3   & 1.24    & 0.0 & $10.8\pm1.3$\\
  2MASS J03480772-6022270 & T7.0  & 2.85 & 7 & 6     & 2.5       & 1.6   & 0.42    & 1.9 & $2.4\pm0.5$\\
  \hline
  \multicolumn{10}{c}{Variables with $5\% < $ $p$-value~$\leq~10\%$ and $\tilde{\eta}\ge1.0$}\\ 
  \hline
  DENIS J0205.4-1159 & L5.5    & 3.22 & 5 & 5     & 2.0 & 1.5 & 0.50 & 7.9   &$2.0\pm0.6$\\
  2MASS J09310955+0327331 & L7.5  & 3.21 & 3 & 10    & 1.8       & 1.0   & 1.55    & 6.1 & $7.6\pm1.9$\\
  2MASS J12171110-0311131 & T7.5    & 2.78 & 6 & 5     & 1.9 & 1.2 & 0.93 & 9.3   &$4.2\pm1.1$\\
\hline
\end{tabular}
\\Notes: $^*$ These peak-to-trough amplitudes are calculated as the difference between the minimum and maximum points in the light curve. In some cases, these might represent the lower limit of the true amplitude, especially for brown dwarfs which exhibit variability on longer time scales.
}
\end{table*}

%__________________________________________________________________

%__________________________________________________________________

\subsection{Identifying variables}
\label{IdentVar}
The significance of the variations were assessed in comparison to two criteria. For the first assessment, the final target light curve was compared against a flat line using the reduced robust median statistic ($\tilde{\eta}$) \citep{enoch03}. The definition of $\tilde{\eta}$ is expressed as 
\begin{equation}
\label{equation1}
\tilde{\eta}= \frac{1}{d}\sum\limits_{i=1}^N \left | \frac{\Delta F_i - {\rm{median}} (\Delta F)}{\sigma_i} \right |
\end{equation}

\noindent where $d$ defines the number of free parameters and $\sigma_i$, the uncertainty on each photometric measurement in the final target light curve.

For the second assessment, the reduced chi squared ($\chi^2_\nu$) value for each target light curve was calculated relative to the master reference light curve. The definition of $\chi^2_\nu$ is expressed as 
\begin{equation}
\label{equation2}
\chi^2_\nu = \frac{1}{\nu}\sum\limits_{i=1}^N \frac{(O_i - E_i)^2}{\sigma_i^2} 
\end{equation}

\noindent where $\nu$ is the degrees of freedom, $O_i$ is the final target light curve, $E_i$ is the master reference light curve and $\sigma_i$ is the uncertainty on the final target light curve and master reference light curve added in quadrature.

Astrophysical variability was better determined calculating $\chi^2_\nu$ relative to the master reference light curve instead of a straight line, which was more prone to classifying variable conditions over intrinsic variability. We make use of the $\chi^2_\nu$ to estimate the cumulative distribution function and thus the $p$-value for each final target light curve. The $p$-value is the probability that the final target light curve is the same as ($p$-value~$>$~10\%) or different from ($p$-value~$\leq$~10\%) the master reference light curve. In Figure~\ref{pval}, we plot the histogram of the calculated $p$-values for the full sample, and the large number of objects in the first bin gives an indication of the variables in the survey. The first bin contains 16 objects with $p$-value~$\leq~5\%$, and the level of false positives expected with an equivalent $p$-value is 3 to 4 objects (5\%) for a sample of 69 targets. For the identification of variables, both $p$-value and an $\tilde{\eta}$ thresholds were applied. The number of targets with $p$-value$\leq5\%$ but $\tilde{\eta}>1$ was two, which is similar to the level of expected false positives. The excess of targets in the last two bins of Figure~\ref{pval} suggests that the uncertainties for the sample are conservatively estimated.

The $p$-value is the probability, under the assumption that we detect no variability (our null hypothesis), of observing variability greater or equal to what was observed in the master reference light curve. The survey has 39 targets satisfying the criterion of $\tilde{\eta}\ge1$ and 16 targets with $p$-value~$\leq~5\%$. Objects classified as variable in this study satisfied two criteria, defined by $p$-value~$\leq~5\%$ and $\tilde{\eta}\ge1$, and they are listed in Table~\ref{tabl:variables}. Candidate variables with a less restrictive $p$-value~$\leq~10\%$ and $\tilde{\eta}\ge1$ are also listed in Table~\ref{tabl:variables}.

The min-to-max amplitudes of the variable objects were calculated as the difference between the highest and the lowest point in the light curve, using the uncertainties on these two points to calculate the uncertainty on the amplitude. Using this method to calculate the amplitude is dependent on how the data is binned, with the unbinned data showing larger amplitude variations. The amplitudes listed in Table~\ref{tabl:variables} use the more conservative estimate from binned data. For objects with periods larger than the duration of observations, the amplitude is likely an underestimate, as the entire period is not observed. An example of a target with a known variable period exceeding the observation timescale is 2M2139, and the reported amplitude in this study is lower than longer timescale results \citep{radigan12}. Due to the limited duration and cadence of the observations, it is not possible to measure the periods of the variables in the BAM study.

%__________________________________________________________________

\section{Results of the BAM survey}
\label{results}
%__________________________________________________________________
\begin{figure*}
\centering
\includegraphics[width=\textwidth]{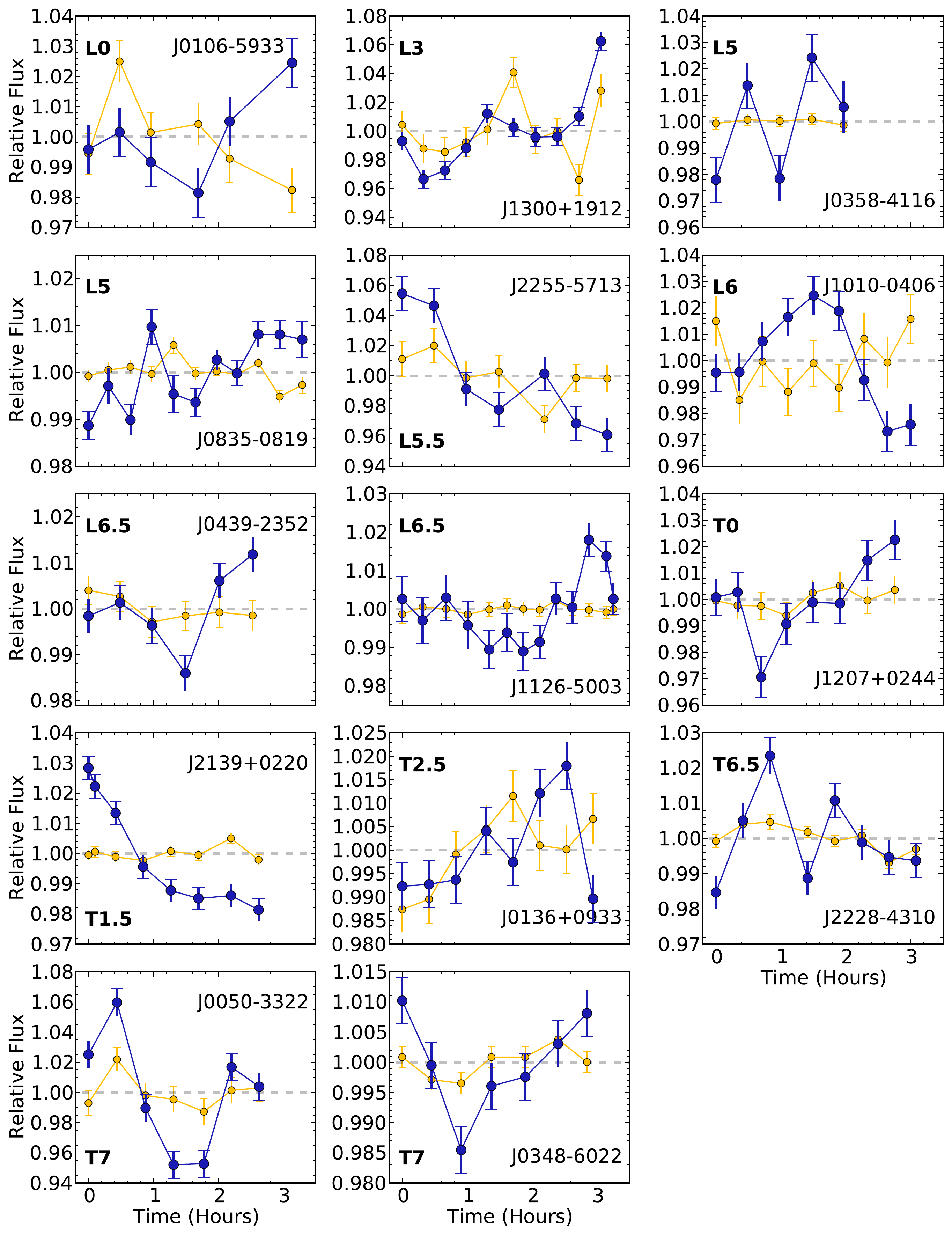}
 \caption{Final target light curves of the 14 variable objects (blue points) with a $p$-value~$\leq~5\%$ and $\tilde{\eta}\ge1.0$ together with the master reference light curves (yellow points). The uncertainties on the variable light curve incorporates the uncertainties in the master reference light curve.} 
  \label{lcs}
\end{figure*}
%__________________________________________________________________
%__________________________________________________________________

\begin{figure*}
\centering
\includegraphics[width=\textwidth]{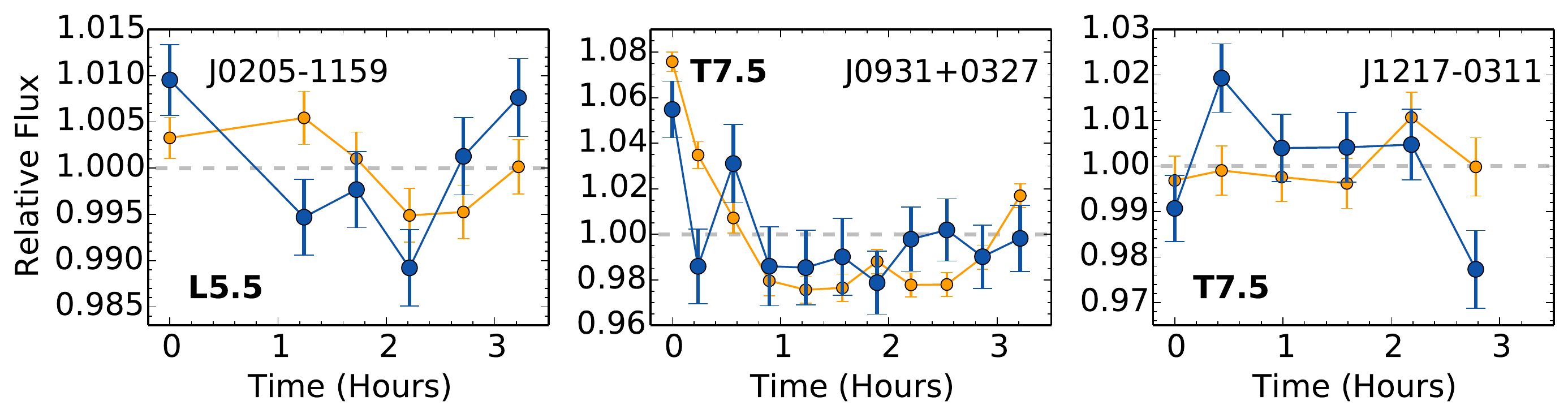}
 \caption{Final target light curves of the candidate variables (larger blue points) with $5\% < p$-value $\leq 10\%$ and $\tilde{\eta}\ge1.0$ together with the master reference light curves (yellow smaller points).} 
  \label{lcs2}
\end{figure*}

%__________________________________________________________________
%__________________________________________________________________
\begin{figure*}
\centering
\includegraphics[width=\textwidth]{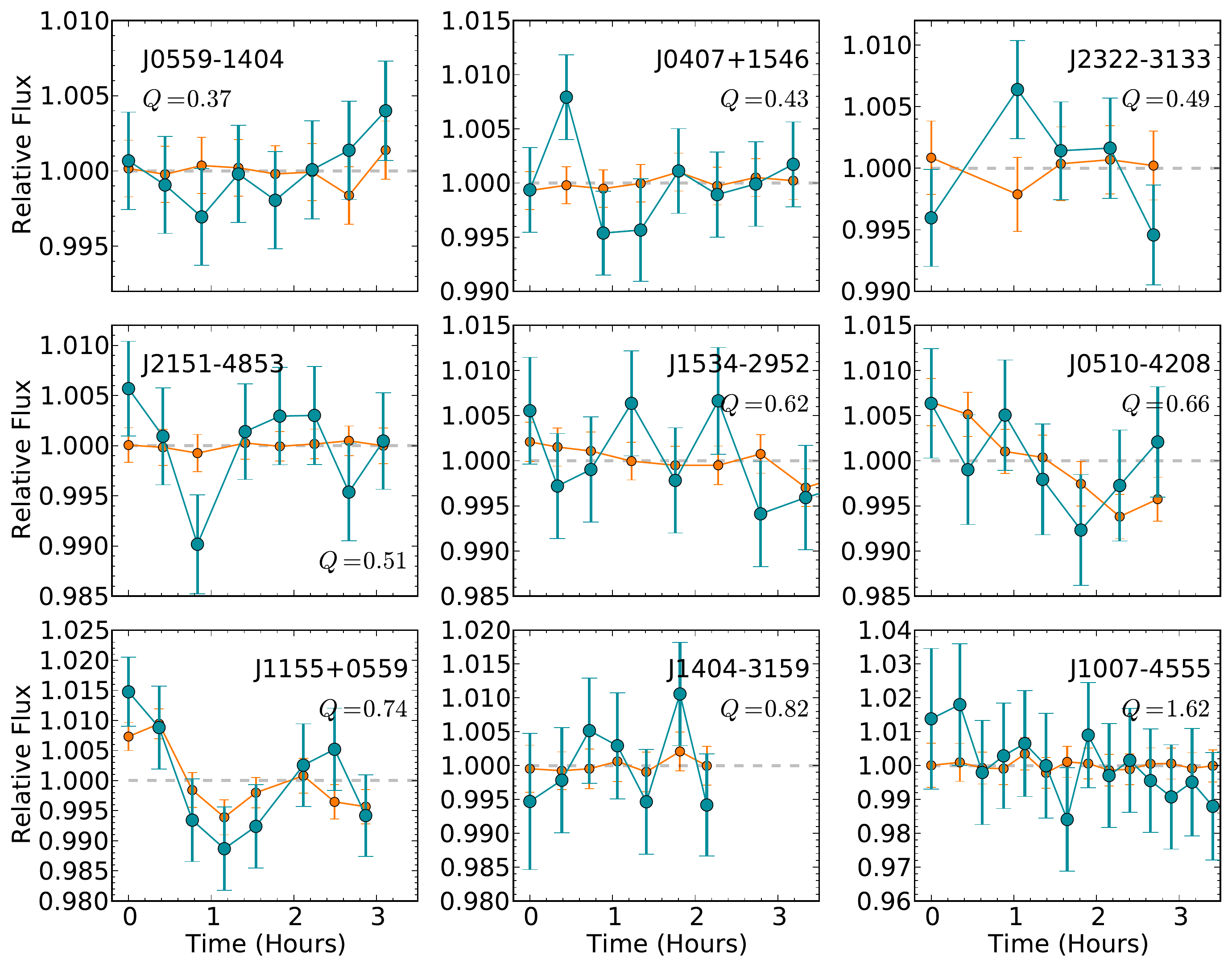}
 \caption{Final target light curves of a subset of constant objects in this survey (larger blue points) with the master reference light curves (yellow smaller points). The target photometric uncertainty decreases from top to bottom and includes the light curve with the best photometry (top left) and light curve with the worst photometry (bottom right).} 
  \label{lcs_constant}
\end{figure*}
%__________________________________________________________________

The primary result from the BAM survey is the identification of a set of 14 variable brown dwarfs with $p$-value~$\leq~5\%$ and a further three candidate variables with $5\%<p$-value $\leq10\%$ (see \S \ref{this-study}). For the remaining analysis, we only consider the $p$-value~$\leq~5\%$ variables. The BAM variables appear to show two morphologies. The first type of light curve shows pure sinusoidal trends, akin to 2M2139 \citep{radigan12}. Variables like 2M0050, 2M0348, 2M1010, and possibly 2M2255 appear to have sinusoidal light curves. The second group consists of targets that appear to be display multi-component variations in their light curves akin to SIMP0136 \citep{artigau09, metchev13}. SIMP0136 shows remarkable evolution in its features over multiple epochs, possibly caused by rapidly varying cloud features \citep{metchev13}. Objects with light curves similar to SIMP0136 object, such as 2M0106, 2M0439, 2M0835, 2M1126, 2M1207, 2M1300 are interesting for future follow up, to confirm whether or not they also show rapidly evolving light curves. Finally, the light curve of 2M0358 may be a fast rotating variable, since 2M0358 appears to oscillate through more than one cycle within the limited timespan of the BAM monitoring. The light curve of 2M2228 shows similar short time scale variations as 2M0358 and was previously found to be variable with a period of P$=1.43^{+0.16}_{-0.15}$~hours by \cite{clarke08}. Final target light curves for the 14 BAM variables and the associated comparison master reference light curves are shown in Figure~\ref{lcs}. Similar plots for the candidate variables are given in Figure~\ref{lcs2}. The amplitudes and $p$-values are reported in Table~\ref{tabl:variables}.

Representative light curves of nine constant targets with a range of photometric qualities are shown in Figure~\ref{lcs_constant}. These light curves show the full range of the data quality for brown dwarfs of similar brightness to the variables identified in the study. The constant light curves are not all flat, however their variations are not statistically distinct from their associated master reference. The constant targets do not satisfy the two separate criteria used to identify the variables (specified in section \ref{IdentVar}) which require the final target light curve to be distinct from the master reference light curve ($p$-value) and a flat line ($\tilde{\eta}$). The $p$-values for constant sources are given in Table \ref{tabl:constants}. 

Since the $p$-value~$\leq~5\%$ cutoff is a statistical measure, there remains a likelihood of a contamination level of 3 -- 4 false variables that are statistical fluctuations, $\sim5$\% of the entire sample. Continued monitoring of the variables should help identify false positives.

\subsection{Comparison of variables with previous studies}
This $J_{\rm s}$-band SofI program is the largest uniform monitoring survey conducted in the near-IR. Several previous surveys have targeted smaller sample sets \citep[e.g.][]{enoch03, koen04b, koen05b, clarke08, khandrika13, girardin13, buenzli13} or searched in different wavelengths such as $I$-band \citep[e.g.][]{gelino02, koen04a, koen13}. Apart from results of the study by \cite{koen13}, previous surveys have typically targeted fewer than $\sim25$ objects and detected variability frequencies of $\sim30$\% in their sample sets, with a significant amount of overlap in the target samples used in different studies \citep{khandrika13}. The BAM sample was designed to uniformly cover the L-T spectral range (see Figure~\ref{spectral_type}) and includes 35 brown dwarfs that have not been previously monitored in different surveys. There are nine new BAM variables, six of which have not been previously monitored for variability -- 2M0050, 2M0106, 2M0358, 2M1010, 2M1207 and 2M2255. The survey has three variables that were previously found to be constant, and found nine brown dwarfs previously classified as variable to be constant. Finally, there are five variables that were found to vary in the literature and in this BAM study. A synopsis of the variables in the BAM and previous surveys is presented in Table~\ref{tabl:variable_sample}. These objects are used to investigate the persistence of variability in section \ref{Persistence}. Table~\ref{tabl:LitConst} presents the constant brown dwarf sample in the BAM study. These are targets that were monitored in previous surveys and were found to be constant in the literature and in this study. In the following two subsections, we compare our results with literature measurements for the variables identified in this sample and in previous work.

\subsubsection{BAM Variables}
\label{this-study}
In Table~\ref{tabl:variable_sample}, we present information for all the targets that were considered variable, either in this BAM study or in the literature. Three of the BAM variables -- 2M0348, 2M0439 and 2M1126 -- were identified as constant brown dwarfs in prior surveys but appear to be variable in this survey. A further five brown dwarfs -- SIMP0136, 2M0835, 2M1300, 2M2139 and 2M2228 -- were confirmed to be variable both in this study and in the literature. Of these five, SIMP0136, 2M2139 and 2M2228 were previously found to vary in the near-IR, similar to this study. The other two variables 2M0835 and 2M1300 were originally measured to vary in the $I_{\rm c}$ band, and also display multi-component variations at near-IR wavelengths. Amongst the known variables with measured periods (from previous studies), only 2M2139 and 2M1300 have periods longer than the duration of the BAM monitoring data ($>7$~hours, and 238~hours, respectively). The latter period is much longer than the expected rotation for a brown dwarf \citep{zapatero06}, which indicates that the periodic feature might not be related to the rotation of the brown dwarf \citep{gelino02}. The three remaining targets with previously measured periods - SIMP0136, 2M0835 and 2M2228 - were monitored in this study with a time span greater than one of their rotational periods. All three objects that we monitored over an entire period had amplitudes consistent with what has been previously published.

\subsubsection{Variables in previous studies not confirmed with BAM}
\label{prev-study}

There are nine targets from the BAM sample that have been previously reported as variable, but were found to be constant in this survey. These sources are listed in Table \ref{tabl:variable_sample}, with a summary of the previous results pertaining to variability, including the observation wavelength and any notes on the amplitudes and timescales of the variations in brightness. One of these nine variables from literature -- 2M0228 -- has only been monitored in the optical. The remaining eight variables exhibited modulations in the near-IR. 2M0559, 2M0624, DENIS0817 and 2M1624 were found to have small amplitude variations in the \citet{buenzli13} survey carried out using the $HST$ grism data. In the $HST$ survey, 2M1624 showed variability in the water band (1.35-1.44$\mu m$) but was found to be constant at $J$-band wavelengths. Similar to other ground based surveys that found some of these targets constant, this BAM survey likely does not have the photometric sensitivity necessary to confirm the $HST$ variables, nor is it possible to monitor the water bands from ground. Another four targets -- SDSS0423, 2M0939, 2M1534 and 2M2331 -- also appear constant in the data. The photometric uncertainties on SDSS0423 and 2M2331 are too large to confirm their lower amplitude variability of $\sim0.8$\% and $\sim1.2$\%, respectively. Despite 2M0939 having been observed as a variable in the $K$' band with an amplitude of 3.1\% \citep{khandrika13}, we are unable to confirm any variability in $J_s$ with the BAM observations. 2M1534 was detected to vary in the $JHK_{\rm s}$ bands initially in \citep{koen04b}, but was constant in a later epoch \citep{koen05b}. \cite{koen13} further discounts the likelihood of detecting short period variability in 2M1534, but maintains that the target likely varies on the timescale of a few days. The reported amplitudes in the $H$-band and $K$-band are below the detection threshold in the data for this target. 

%__________________________________________________________________

\section{Discussion}
\label{discussion}

\subsection{The sensitivity of the BAM survey}  
\begin{figure}
\centering
\includegraphics[width=84mm]{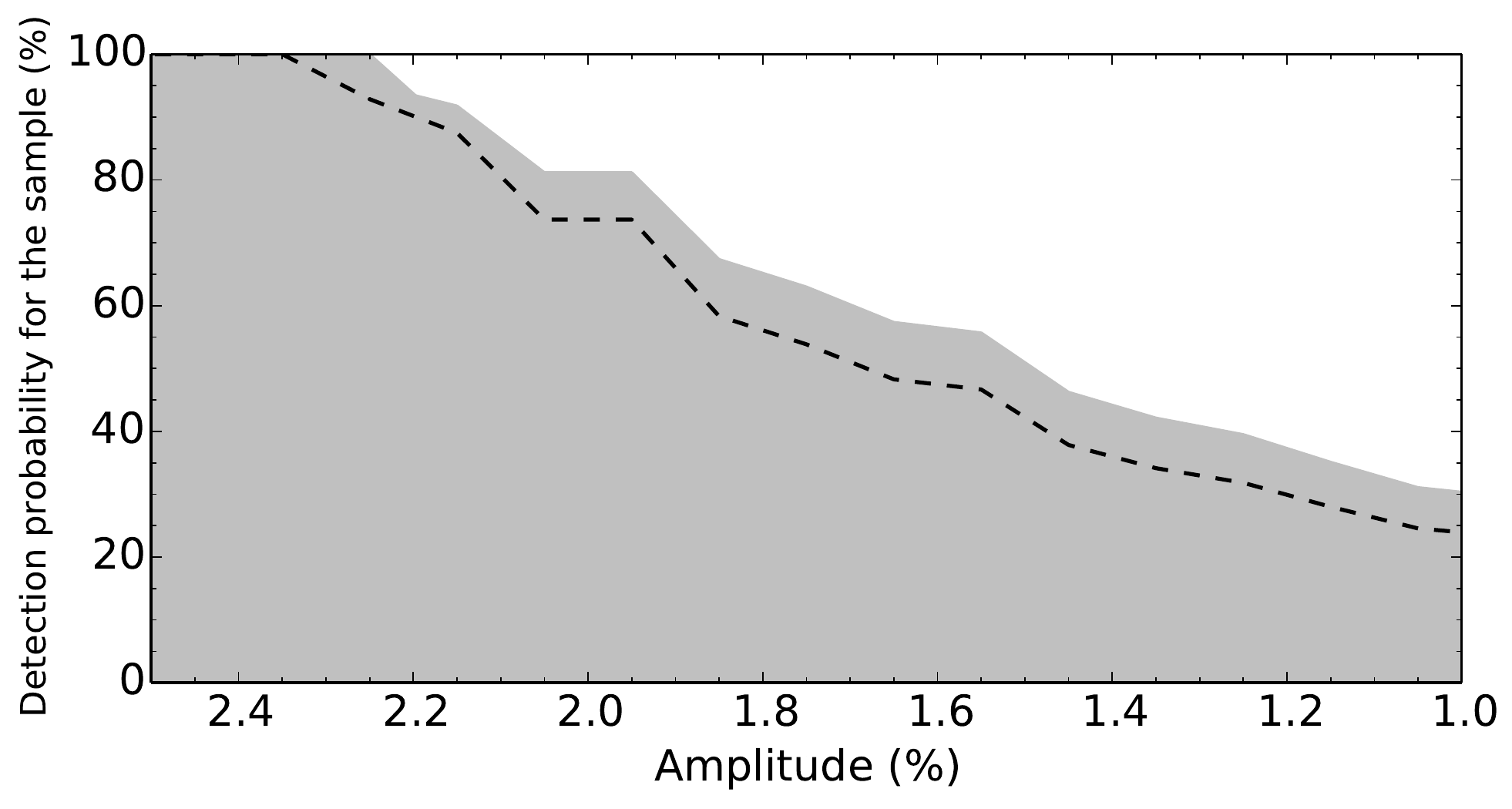}
 \caption{Proportion of the survey sensitive to variability as a function of peak-to-trough amplitudes for different detection thresholds. The dashed line represents the fraction of objects with a photometric accuracy good enough to have allowed for the detection of variability. The shaded area represents the region of sensitivity with the upper binomial errors and amplitude uncertainties added to the variability fraction.}
  \label{sensitivity}
\end{figure}

\begin{figure}
\centering
\includegraphics[width=84mm]{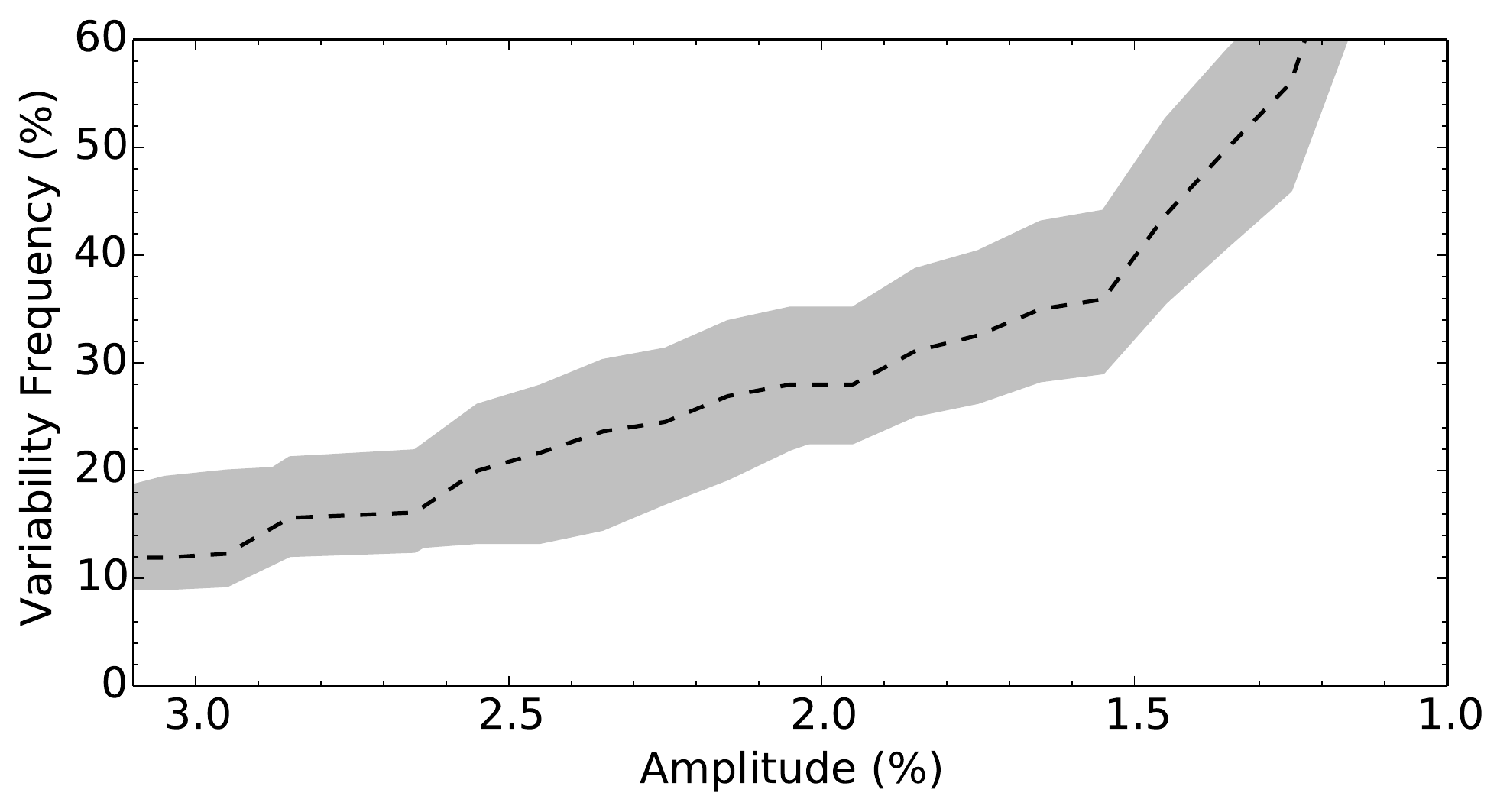}
 \caption{Variability frequency as a function of amplitude (dashed line) with the binomial errors and amplitude uncertainties added to the variability fraction (shaded area).}
  \label{var_fraction}
\end{figure}

To obtain an estimate of the variability frequency for brown dwarfs across spectral types, it is essential to quantify the sensitivity of the data to detecting different amplitudes of variability. We estimate the sensitivity to variables of a certain amplitude as three times the target photometric uncertainty of each final target light curve. This places a limit on the minimum amplitude required for a detection above a certain statistical significance threshold. The proportion of the sample that is sensitive to a given variability amplitude is shown as a function of amplitude in Figure~\ref{sensitivity}. As shown in Figure~\ref{sensitivity}, the BAM survey is capable of detecting any object in the sample showing a peak-to-trough amplitude $\ge2.3\%$ during the duration of the observations. The detection probability continues to decrease with decreasing amplitude with a sensitivity of 50\% occurring for variables with a $\sim1.7$\% amplitude. Given that the full BAM sample is sensitive to variables with amplitudes $\ge2.3\%$, Table~\ref{tabl:AmpLimits} quantifies the frequency of variability for different subsets of spectral types using an amplitude cutoff of 2.3\% and $p$-value $\le0.05$; this level includes all but one BAM $p\le0.05$ variable. Figure~\ref{var_fraction} shows how the variability frequency (considering all spectral types) varies as a function of amplitude to account for the declining proportion of the sample that is sensitive to lower amplitude variables.

To calculate the uncertainty on the variability frequency we use the binomial distribution

%__________________________________________________________________
\begin{table*}
\tiny{\centering
\caption{Limits on constant targets in this survey.}
\label{tabl:constants}
\begin{tabular}{lcccccccc}
\hline
Object &   Spectral Type &  Obs. Dur. [hours] & Refs. & DOF &  $\chi^2_\nu$ & $\tilde{\eta}$ & $Q$ (\%) & $p$-value (\%)\\
\hline
\hline
  2MASS J00165953-4056541 & L3.5 & 3.43 & 6 & 8 & 0.9 & 0.8 & 0.54 & 51.1\\
  2MASS J00184613-6356122 & L2 & 3.43 & 7 & 8 & 1.1 & 1.0 & 0.42 & 35.4\\
  2MASS J00345157+0523050 & T6.5 & 2.98 & 7 & 7 & 0.9 & 1.1 & 0.53 & 49.1\\
  2MASS J01282664-5545343 & L1 & 2.66 & 5 & 5 & 0.5 & 2.1 & 0.61 & 77.7\\
  2MASS J02284355-6325052 & L0 & 2.84 & 3 & 10 & 0.8 & 1.0 & 0.42 & 62.7\\
  2MASS J02572581-3105523 & L8 & 3.33 & 7 & 5 & 0.3 & 1.2 & 0.48 & 90.2\\
  2MASS J03185403-3421292 & L7 & 1.97 & 5 & 4 & 1.9 & 1.8 & 0.99 & 10.1\\
  2MASS J03400942-6724051 & L7 & 2.61 & 6 & 5 & 0.2 & 1.1 & 0.75 & 96.4\\
  2MASS J04070752+1546457 & L3.5 & 3.19 & 7 & 7 & 0.7 & 0.7 & 0.43 & 68.4\\
  2MASS J04151954-0935066 & T8 & 3.29 & 4 & 7 & 0.3 & 0.6 & 0.83 & 95.8\\
  SDSS J042348.56-041403.5 & T0 & 2.62 & 8 & 5 & 0.4 & 0.7 & 0.51 & 86.2\\
  2MASS J04455387-3048204 & L2 & 4.55 & 7 & 9 & 0.7 & 0.7 & 0.36 & 68.7\\
  2MASS J05103520-4208140 & T5 & 2.74 & 6 & 6 & 0.5 & 0.6 & 0.66 & 84.2\\
  2MASS J05160945-0445499 & T5.5 & 3.11 & 4 & 7 & 1.1 & 1.1 & 0.88 & 38.5\\
  2MASS J05233822-1403022 & L5 & 4.56 & 6 & 9 & 0.7 & 1.0 & 0.99 & 66.8\\
  2MASS J05591914-1404488 & T4.5 & 3.11 & 6 & 7 & 0.3 & 0.5 & 0.37 & 96.3\\
  2MASS J06244595-4521548 & L5 & 3.26 & 5 & 6 & 1.2 & 1.0 & 0.49 & 28.5\\
  2MASS J07290002-3954043 & T8 & 3.37 & 8 & 10 & 0.6 & 0.6 & 0.9 & 84.3\\
  DENIS J081730.0-615520 & T6 & 3.48 & 8 & 13 & 0.8 & 0.7 & 0.68 & 64.5\\
  2MASS J09153413+0422045 & L7 & 3.33 & 6 & 7 & 0.9 & 0.8 & 0.55 & 50.1\\
  2MASS J09393548-2448279 & T8 & 3.06 & 8 & 9 & 1.0 & 0.8 & 0.98 & 46.5\\
  2MASS J09490860-1545485 & T2 & 3.16 & 8 & 6 & 1.7 & 1.3 & 0.67 & 11.0\\
  2MASS J10043929-3335189 & L4 & 3.14 & 8 & 11 & 1.4 & 1.0 & 0.99 & 17.6\\
  2MASS J10073369-4555147 & T5 & 3.41 & 8 & 13 & 0.3 & 0.5 & 1.62 & 99.5\\
  2MASS J10210969-0304197 & T3 & 3.17 & 7 & 6 & 0.7 & 0.6 & 0.88 & 64.1\\
  2MASS J11145133-2618235 & T7.5 & 2.90 & 5 & 5 & 0.4 & 1.0 & 1.0 & 86.3\\
  2MASS J11555389+0559577 & L7.5 & 2.87 & 5 & 7 & 0.6 & 1.2 & 0.74 & 79.3\\
  2MASS J12255432-2739466 & T6 & 2.85 & 7 & 5 & 0.4 & 0.3 & 0.67 & 86.9\\
  2MASS J12281523-1547342 & L6 & 3.39 & 7 & 10 & 1.0 & 1.1 & 0.5 & 42.5\\
  2MASS J12314753+0847331 & T5.5 & 2.56 & 5 & 5 & 0.3 & 1.6 & 1.63 & 89.6\\
  2MASS J12545393-0122474 & T2 & 3.17 & 7 & 9 & 1.0 & 1.1 & 0.51 & 45.2\\
  2MASS J13262981-0038314 & L5.5 & 2.80 & 4 & 7 & 0.9 & 0.9 & 1.17 & 50.8\\
  2MASS J14044941-3159329 & T2.5 & 3.01 & 8 & 8.5 & 0.4 & 0.6 & 0.82 & 89.4\\
  2MASS J15074769-1627386 & L5.5 & 4.16 & 7 & 15 & 0.4 & 0.6 & 1.21 & 98.5\\
  SDSS J151114.66+060742.9 & T.0 & 3.85 & 7 & 11 & 0.8 & 0.6 & 0.8 & 64.1\\
  2MASS J15210327+0131426 & T2 & 6.37 & 7 & 12 & 0.5 & 0.6 & 0.9 & 93.6\\
  2MASS J15344984-2952274 & T5.5 & 3.88 & 8 & 8 & 0.5 & 0.6 & 0.62 & 85.3\\
  2MASS J15462718-3325111 & T5.5 & 3.50 & 4 & 7 & 0.4 & 0.5 & 1.45 & 93.3\\
  2MASS J15530228+1532369 & T7 & 3.82 & 7 & 8 & 2.7 & 1.0 & 0.72 & 0.7\\
  2MASS J16241436+0029158 & T6 & 3.21 & 8 & 8 & 0.3 & 0.4 & 0.66 & 95.6\\
  2MASS J16322911+1904407 & L8 & 3.89 & 4 & 14 & 1.1 & 1.0 & 1.76 & 36.4\\
  2MASS J18283572-4849046 & T5.5 & 3.06 & 8 & 7 & 0.4 & 0.5 & 0.5 & 88.7\\
  2MASS J19360187-5502322 & L5 & 3.21 & 7 & 6 & 0.4 & 0.5 & 0.4 & 86.9\\
  SDSS J204317.69-155103.4 & L9.0 & 3.03 & 8 & 6 & 1.4 & 1.0 & 1.14 & 22.5\\
  SDSS J204749.61-071818.3 & T0.0 & 2.68 & 8 & 5 & 1.6 & 1.2 & 1.16 & 15.8\\
  2MASS J20523515-1609308 & T1 & 3.08 & 8 & 7 & 0.8 & 0.8 & 0.7 & 62.9\\
  2MASS J21513839-4853542 & T4 & 3.08 & 8 & 7 & 0.8 & 0.7 & 0.51 & 63.0\\
  2MASS J22521073-1730134 & L7.5 & 2.34 & 5 & 5 & 0.8 & 1.3 & 0.65 & 58.9\\
  ULAS J232123.79+135454.9 & T7.5 & 2.94 & 8 & 7 & 0.3 & 0.6 & 0.9 & 94.9\\
  2MASS J23224684-3133231 & L0 & 2.69 & 6 & 4 & 1.1 & 0.9 & 0.49 & 37.8\\
  2MASS J23312378-4718274 & T5 & 2.63 & 5 & 6 & 1.7 & 1.1 & 1.13 & 12.0\\
  2MASS J23565477-1553111 & T6 & 2.98 & 5 & 7 & 3.5 & 0.6 & 0.64 & 0.1\\
\hline
\end{tabular}
%\\Notes: $^a$ Discontinuous. $^b$ Deviant data point. $^c$ Very noisy master reference light curve. 
}
\end{table*}

%__________________________________________________________________

\begin{table*}
\tiny{\centering
\caption{Summary of variable sources.}
\label{tabl:variable_sample}
\begin{tabular}{lcccl}
\hline
\multicolumn{1}{c}{Target Name}    & Band    &    Variable/Constant		&	References & \multicolumn{1}{c}{Notes} \\
\hline
\hline

  \multicolumn{5}{c}{New variables from this study with no prior observations} \\
  \hline
  2MASS J00501994-3322402 & $J_{\rm s}$ & V &  &  \\
  2MASS J01062285-5933185 & $J_{\rm s}$ & V &  &  \\
  2MASS J03582255-4116060 & $J_{\rm s}$ & V &  &  \\
  2MASS J09310955+0327331 & $J_{\rm s}$ & V &  & Candidate Variable ($p$-val$ = 6.1$\%) \\
  2MASS J10101480-0406499 & $J_{\rm s}$ & V &  &  \\
  2MASS J12074717+0244249 & $J_{\rm s}$ & V &  &  \\
  2MASS J22551861-5713056 & $J_{\rm s}$ & V &  &  \\

  \hline
  \multicolumn{5}{c}{New variables previously categorised as constant} \\
  \hline
  DENIS J0205.4-1159 & $I_{\rm c}$ & C & K13 & \\
  Candidate Variable ($p$-val$ = 7.9$\%) & $K_{\rm s}$ & C & E03 & $<3$\%\\
   & 8.46 GHz & C & B06 & $< 30 \mu\rm{Jy}$\\
  2MASS J03480772-6022270 & $J$  & C & C08 & $<10$~mmag, periodic \\
  2MASS J04390101-2353083 & $I_{\rm c}$ & C & K13 & \\
   & $I_{\rm c}$ & C & K05 & \\
   & 8.46 GHz & C & B06 & $<$42$\mu$Jy\\
  2MASS J11263991-5003550 & $I_{\rm c}$ & C & K13 & possibly periodic \\
  \hline
  \multicolumn{5}{c}{Literature variables confirmed as variable in this study} \\
  \hline
  SIMP J013656.5+093347.3 & $JK$ & V & A09, Ap13 & $\Delta J=4.5\%$, P = 2.39 $\pm$ 0.05 hr\\
  2MASS J08354256-0819237 & $I_{\rm c}$ & V & K04, K13 & $\Delta I_{\rm c}$=10-16~mmag, P=3.1~hr    \\
        & 8.46 GHz & C & B06 & $<30\mu\rm{Jy}$\\
  2MASS J12171110-0311131 & $J$ & V & A03 & $\Delta J = 0.176 \pm 0.013$~mag \\
  Candidate Variable (p-val$ = 9.3$\%) & 8.46 GHz & C & B06 & $<111\mu\rm{Jy}$\\
        & & & Z06 &  \\
  2MASS J13004255+1912354 & $I_{\rm c}$ & C & K13 & \\
        & $I$ & V & G02 & P=238$\pm$9hr\\
        & 8.46 GHz & C & B06 & $<87\mu\rm{Jy}$\\
		&  $J+K$' & C & Kh13 & $J <$ 1.1\%, $K$' $<$ 1.7\%\\
		&  $I_{\rm c}$ & C & K05 & \\
		&  $JHK_{\rm s}$ & C & KMM04 &\\
  2MASS J21392676+0220226 & $JHK_{\rm s}$ & V & R12, Ap13 & $\Delta$($J$,$H$,$K$)=(0.3, 0.18, 0.17)~mag, P = 7.721$\pm$0.005~hr \\
        & $J+K$' & V & Kh13 & $\Delta J=6.7\%$, C at $K$' $<8$\%\\
  2MASS J22282889-4310262 & $J$ & V & C08, Bu12 & $\Delta J=15.4 \pm 1.4$~mmag, P=1.43$\pm$0.16hr\\
        & 8.46 GHz & C & B06 & $<30\mu\rm{Jy}$\\

  \hline
  \multicolumn{5}{c}{Objects with reported IR variability measured as constants in this study} \\
  \hline
  SDSS J042348.56-041403.5 & $I_{\rm c}$ & C & K13 & \\
   & $K_{\rm s}$ & likely V & E03 & $0.3\pm0.18$~mag, P$=1.39$~--~$1.62$~hr\\
   & $J$ & V & C08 & $8.0\pm0.8$~mmag, P$=2\pm0.4$hr\\
   & $JHK_{\rm s}$ & C & KTTK05 & $J$ $<15$~mmag, $H< 11$~mmag, $K <2$~mmag\\
   & 8.46 GHz & C & B06 & $<42\mu\rm{Jy}$\\
  2MASS J05591914-1404488  & HST G141 grism & V & Bu13 & \\
   & $I_{\rm c}$ & C & K13 & \\
   & $K_{\rm s}$ & C & E03 & $<7$\%\\
   & $J$ &  C08 & C & $<5$~mmag\\
   & $I_{\rm c}$ & C & K04 & \\
   & 8.46 GHz &  B06 & $<27\mu\rm{Jy}$\\
  2MASS J06244595-4521548  & HST G141 grism & V & Bu13 & \\
  DENIS J081730.0-615520  & HST G141 grism & V & Bu13 & \\
  2MASS J09393548-2448279 & $K$' & V & Kh13 & 0.31~mag\\
   & $J$ & C &  & $<0.141$~mag\\
  2MASS J15344984-2952274 & $I_{\rm c}$ & C & K13 & \\
   & $I_{\rm c}$ & C & K05 & \\
   & $JHK_{\rm s}$ & C & KTTK05 & $J <10$~mmag, $H <11$~mmag, $K <$18~mmag\\
   & $JHK_{\rm s}$ & V & KMM04 & $H$ $4$~mmag, $K$ 7~mmag, P$=0.96$~hr\\
   & 8.46 GHz & C & B06 & $<63\mu\rm{Jy}$\\
  2MASS J16241436+0029158 & HST G141 grism & V & Bu13 & Variability detected in water band (1.35-1.44~$\mu m$)\\
   & $JHK_{\rm s}$ & C & KMM04 & \\
   & 8.46 GHz & C & B06 & $<$36$\mu$Jy\\
   2MASS J23312378-4718274 & $J$ & V & C08 & $\Delta J=12.4 \pm 1.3$~mmag, P=2.9$\pm$0.9~hr \\
   
  \hline
  \multicolumn{5}{c}{Objects with reported Optical variability measured as constants in this study} \\
  \hline
  2MASS J02284355-6325052 & $I_{\rm c}$ & V & K13 & \\
\hline
\end{tabular}

References:    \citet{artigau03} [A03], 
     \citet{artigau09} [A09],
     \citet{apai13} [Ap13],
     \citet{berger06} [B06],
     \citet{buenzli12} [Bu12],
     \citet{buenzli13} [Bu13],
     \citet{clarke08} [C08],
     \citet{enoch03} [E03],
     \citet{gelino02} [G02],
     \citet{khandrika13} [Kh13],
     \citet{koen04a} [K04],
     \citet{koen04b} [KMM04],
     \citet{koen05b} [KTTK05],
     \citet{koen05a} [K05],
     \citet{koen13} [K13],
     \citet{radigan12} [R12],
     \citet{zapatero06} [Z06]\\
}
\end{table*}

%__________________________________________________________________

\begin{table}
\tiny{ \centering
\caption{Summary of constant sources.}
\label{tabl:LitConst}
\begin{tabular}{lccl}
\hline
\hline
\multicolumn{1}{c}{Target Name}    & Band    &    References & \multicolumn{1}{c}{Notes} \\
 \hline
  2MASS J02572581-3105523 & $I_{\rm c}$ &  K13 & \\
  2MASS J04070752+1546457 & $JK'$ &  Kh13 & no results\\
  2MASS J04151954-0935066 & 8.46 GHz  & B06 & $<\mu\rm{Jy}$\\
  2MASS J04455387-3048204 & $I_{\rm c}$ & K13 & \\
   & $I_{\rm c}$ &  K04 & \\
   & 8.46 GHz &  B06 & $<66\mu\rm{Jy}$\\
  2MASS J05233822-1403022 & $I_{\rm c}$ & K13 & \\
   & $I_{\rm c}$ &  K05 & \\
   & 8.46 GHz &  B06 & $231\pm14\mu\rm{Jy}$\\
  2MASS J12255432-2739466 & $J$ &  KTTK05 & $<12$~mmag\\ & $H$ & &  $<14$~mmag\\ & $K_{\rm s}$ & &  $<9$~mmag\\
   & $JHK_{\rm s}$ &  KMM04 & \\
  2MASS J12281523-1547342 & $I_{\rm c}$ &  K13 & \\
   & 8.46 GHz &  B06 & $<87\mu\rm{Jy}$\\
  2MASS J12545393-0122474 & $JHK_{\rm s}$ &  KMM04 & \\
   & $J$ & Gi13 & $<5$mmag \\
  2MASS J15074769-1627386 & $I_{\rm c}$ &  K13 & \\
   & $I_{\rm c}$ &  K03 & \\
   & 8.46 GHz &  B06 & $<57\mu\rm{Jy}$\\
  2MASS J1511145+060742 & $J$ &  Kh13 & $<3.3$\%\\  & $K$'& & $<6.1$\%\\
  2MASS J15462718-3325111 & $JHK_{\rm s}$ &  KMM04 & \\
  2MASS J15530228+1532369 & $JHK_{\rm s}$ &  KMM04 & \\
  2MASS J16322911+1904407 & 8.46 GHz &  B06 & $<54\mu$Jy\\
   & HST G141 Grism &  Bu13 & \\
  2MASS J19360187-5502322 & $I_{\rm c}$ &  K13 & \\
  2MASS J23224684-3133231 & $I_{\rm c}$ &  K13 & \\
 \hline
\end{tabular}
References: \citet{berger06} [B06],
     \citet{buenzli13} [Bu13],
     \citet{girardin13} [Gi13],
     \citet{khandrika13} [Kh13],
     \citet{koen04a} [K04],
     \citet{koen03} [K03],
     \citet{koen04b} [KMM04],
     \citet{koen05b} [KTTK05],
     \citet{koen05a} [K05],
     \citet{koen13} [K13].\\
}
\end{table}

%__________________________________________________________________
%__________________________________________________________________
%__________________________________________________________________
%__________________________________________________________________
\begin{figure}
\centering
\includegraphics[width=84mm]{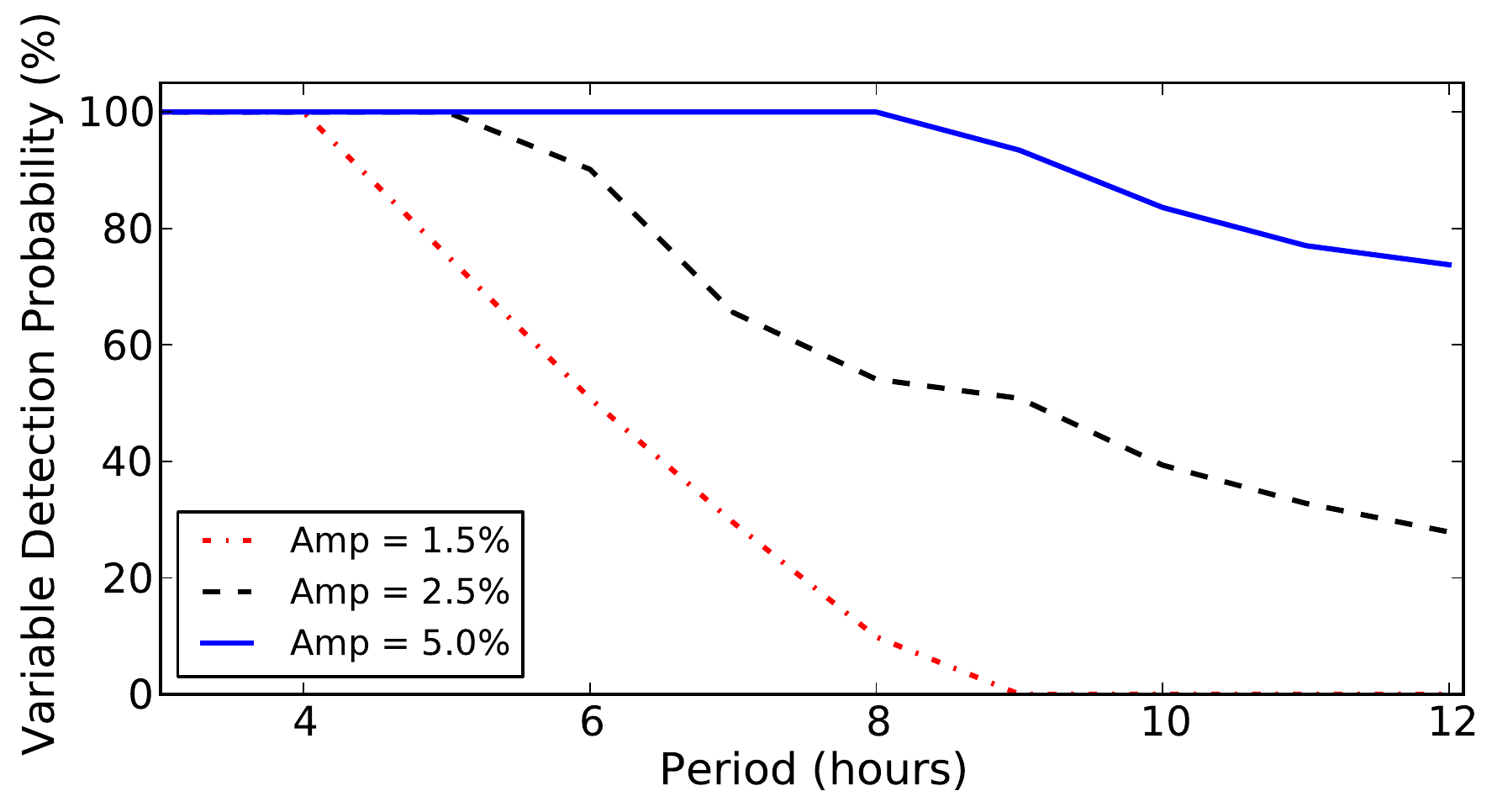}
\caption{Percentage of simulated sinusoidal light curves detected as variable, as a function of period from 1~hour to 12~hours, for three different amplitudes. We used the measured survey median noise of 0.7\%, and each sine curve was sampled at intervals of 15 minutes to imitate the binned data of the survey. Additionally, we stepped through each sine curve at 5 degree phase intervals, to ensure that we sampled the full phase of the variable light curve.}
  \label{per_sensitivity}
\end{figure}
%__________________________________________________________________

\begin{equation}
B(n;N,{\epsilon}_v) = \frac{N!}{n!(N-n)!}{\epsilon}_v^n(1-{\epsilon}_v)^{N-n}.
\label{equation3}
\end{equation}

\noindent where $n$ is the number of variables, $N$ the sample size and ${\epsilon}_v$ the variability frequency. This approach is based on Bayes' theorem under the assumption of a uniform prior based on no a priori knowledge and is ideal for small samples such as is the case for the BAM survey.

The rotation period is another factor that can influence the detectability of a variable signal. In Figure~\ref{per_sensitivity}, we present the results of simulating light curves to test the detection probability of the survey to brown dwarf variables with different periods. We simulated sinusoidal light curves with three different amplitudes, of 1.5\%, 2.5\%, and 5.0\%, and with periods ranging from a minimum of 1~hour to a maximum of 12 hours \citep{zapatero06}. Gaussian noise equal to the median photometric uncertainty of the survey of 0.7\% was added to each light curve. To mimic the binned SofI data, the light curves were sampled at intervals of 15~minutes, and each simulated dataset was divided into groups of 3 hours, similar to the typical duration of the BAM data. For light curves with period longer than 3 hours, we generated multiple datasets, by stepping through the sine curve in steps of 5 degrees of phase and calculating the $p$-value at each phase, ensuring full sampling of the phase. Figure~\ref{per_sensitivity} shows the percentage of simulated light curves that are detected as variable with a $p$-value $\le5\%$. For amplitudes of 5.0\%, periodicities from 3 to 12~hours are easily recovered with a probability of 80 to 100\%, while the required periods decrease to $\sim$6~hours for 2.5\% variables and $\sim$5~hours for 1.5\% variables for detection probabilities in the 80 to 100\% range.

%__________________________________________________________________

\subsection{Frequency and amplitude of variability across spectral types}
The frequency of variables as a function of spectral type is an important topic, since models of brown dwarf atmospheres have suggested that breakup of clouds across the L/T transition may result in both a higher rate of occurrence and a higher amplitude of variability compared to earlier L and later T objects. Amongst the previously known variables, the two largest amplitude variable objects discovered to-date are L/T transition objects - SIMP0136 ($\sim5$\% in $J$-band but with a significant night to night evolution, \citealt{artigau09}) and 2M2139 (as high as 26\% in the $J$-band, \citealt{radigan12}).

%__________________________________________________________________

\begin{table*}
\centering
\caption{Variability frequency.}
\label{tabl:AmpLimits}
\begin{tabular}{ccccc}
\hline
Sample   & Sp. Type & No.~Targets  & No.~Variables$\dagger$  & Freq. (\%)\\
\hline
\hline
  Early-L                & L0-L6            & 23  & 7  & $30^{+11}_{-8}$\\ 
  Late-T                 & T5-T8            & 23  & 3  & $13^{+10}_{-4}$\\
  L/T Transition         & L7-T4            & 23  & 3  & $13^{+10}_{-4}$\\ 
  Outside L/T transition & L0-L6 \& T5-T8   & 46  & 10 & $22^{+7}_{-5}$\\
\hline

\end{tabular}
\\Notes: $^\dagger$ These are the variables with a $p$-value~$\leq~5\%$, and amplitude $\ge2.3\%$.
\end{table*}
%__________________________________________________________________
%__________________________________________________________________
\begin{figure*}
\centering
\includegraphics[width=96mm]{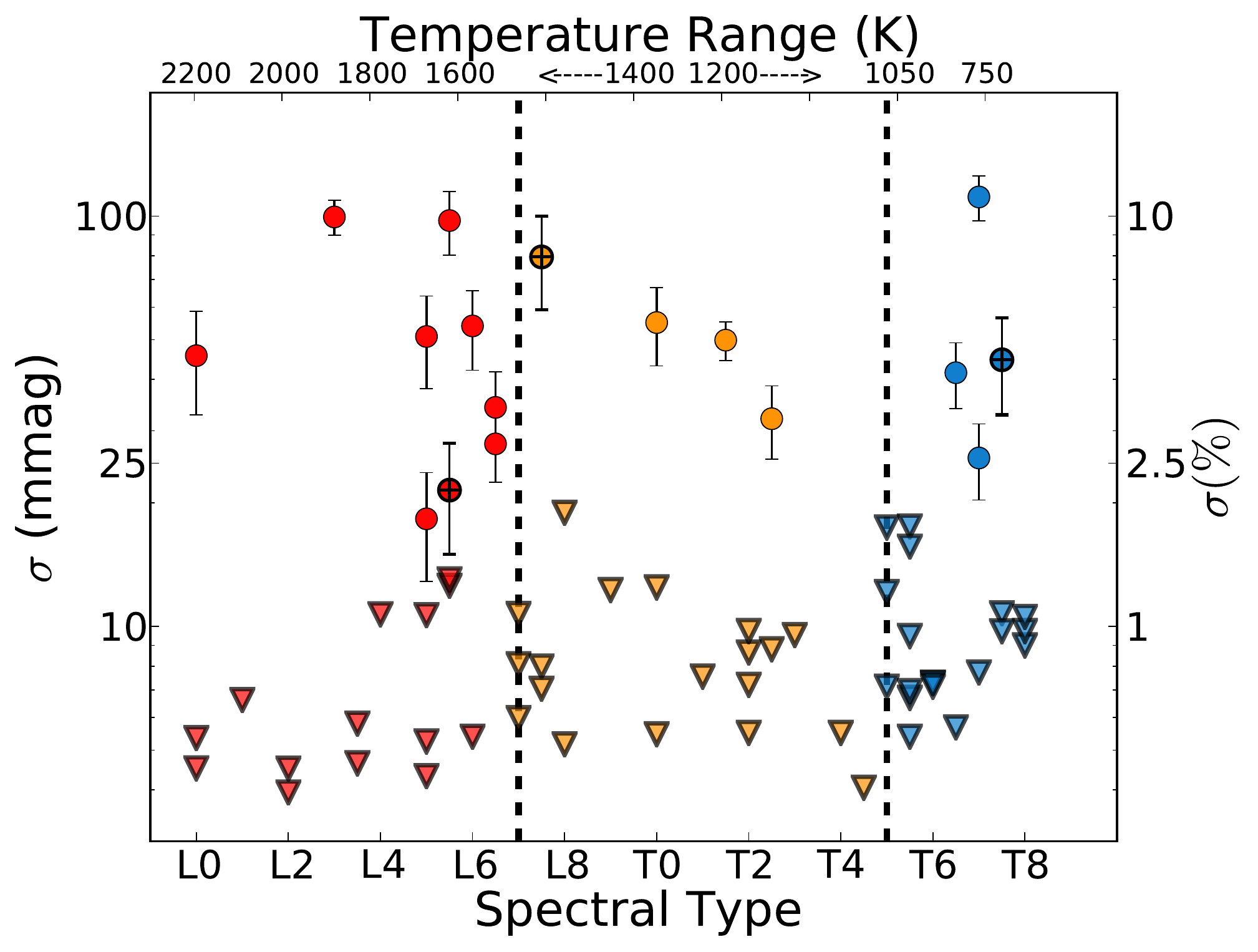}
\includegraphics[width=87mm]{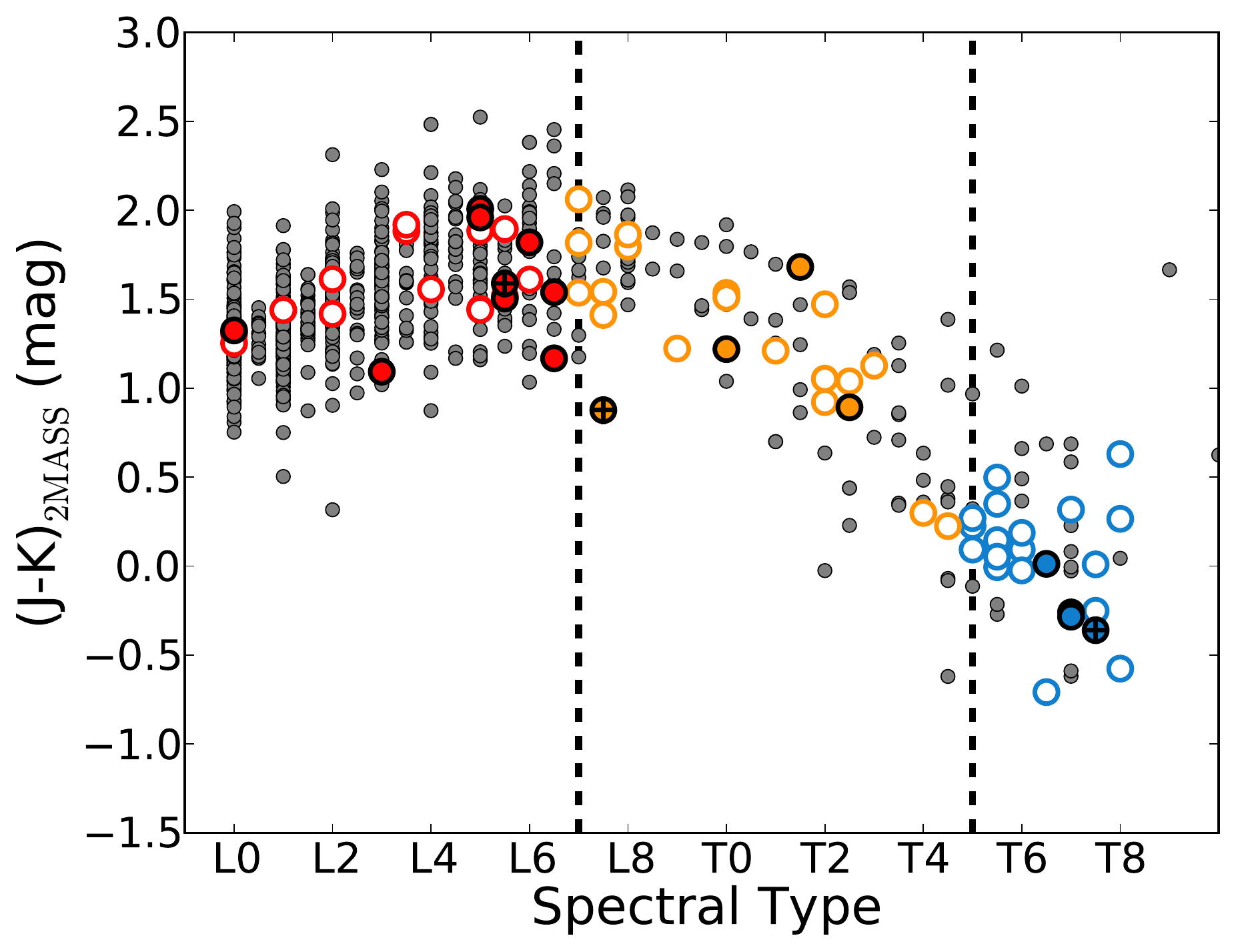}
 \caption{Diagram on the left shows the amplitude of the variables ($p$-value~$\leq~5\%$ -- closed circles and $5\% <$ $p$-value~$\leq~10\%$ -- closed circles with cross) as well as the target photometric uncertainty of the non-varying objects (coloured triangles) across the entire spectral range of the sample. The diagram on the right shows the colour-colour diagram of the entire L through T spectral range with the full sample plotted with open circles, showing the colour spread of the targets. The variables from the BAM sample are overplotted ($p$-value~$\leq~5\%$ -- closed circles and $5\% <$ $p$-value~$\leq~10\%$ -- closed circles with cross). The L/T transition is indicated by the dashed lines} 
  \label{survey}
\end{figure*}
%__________________________________________________________________

As indicated by the variability frequencies reported in Table~\ref{tabl:AmpLimits}, the BAM results show no evidence that the frequency of variables in the L7 to T4 transition region is distinct from the earlier spectral types, the later spectral types, or the combination of all non-transition region brown dwarfs. The variability frequencies in Table~\ref{tabl:AmpLimits} are calculated using the entire sample of targets and an amplitude threshold of $\ge2.3\%$ and $p\le0.05$. Although no statistically significant difference in the variability frequencies for transition brown dwarfs is measured with the BAM survey, the 2.3\% amplitude limit of the analysis would not have detected differences at lower amplitudes, and removing the peak-to-trough amplitude threshold does not change this result. Adjusting the boundaries of the transition region by up to two spectral types does not change the result. Likewise, the amplitudes of the detected variables show no clear trends with spectral type within the capacity of the survey, as shown in Figure~\ref{survey} (left).

The BAM variability frequency is very comparable to estimates for M stars. A variability frequency between $\sim$21--29\% for 19 M-stars was measured in a multi-wavelgenth optical study with the Calar Alto Observatory in Spain \citep{rockenfeller06}. The wavelength of observations for the M-star study was shorter than the BAM survey $J$-band data, and the amplitude of variations is expected to decline for longer wavelengths \citep[e.g.][]{reiners10}.  

In a recent compilation of variability surveys, \cite{khandrika13} reported a variability frequency of $30\pm5$\% based on a collection of different surveys with observations obtained in the optical and near-IR passbands, covering 78 objects in total. Comparison between surveys is difficult as the variability frequency may depend on a variety of different factors, including the target selection criterion and the criteria used to define variability in the targets which usually differs from one survey to the next. Additionally, the observed wavelength may also alter the variability frequency with different wavelength probing different depths in the atmosphere. \cite{koen13} finds a poor overlap between the variables identified with optical and near-IR filters (of the 13 variables already observed in near-IR surveys, 7 were found as constant and 6 as variable in the optical). Because of the uniform sensitivity of this survey, we did not incorporate the results of previous studies into the statistics, presented in Table~\ref{tabl:AmpLimits}. The presence of highly variable objects outside the transition region, may suggest the possibility of both early onset of cloud condensation in the atmospheres of mid-L dwarfs and the emergence of sulfide clouds in mid-T dwarfs \citep{Morley12}. Other physical processes that have been suggested to possibly induce variability in the atmospheres of brown dwarfs include coupling clouds with global atmosphere circulation \citep{showman13,zhang14}, and variability caused by thermal perturbations emitted from deeper layers within the brown dwarf atmosphere \citep{robinson14}.

\subsection{Variability as a function of colour within a spectral type}
The $J-K$ colour of the sample as a function of the spectral type is shown in Figure~\ref{survey} (right). The targets span nearly the full colour spread in early-L, transition and late-T sub sample. The 14 BAM variables and the three candidates are not clustered toward either the red or the blue within any particular spectral type. Previous studies \citep[e.g.][]{khandrika13} have suggested that brown dwarfs with unusual colours (highly red or blue) compared to the median of the spectral type might be indicative of variable cloud cover. We performed a two sample K-S test to determine whether or not the detrended colors of the BAM variables were distinct from the rest of the sample. The maximum difference between the cumulative distributions was 0.18 with a corresponding $p$-value of $\sim75\%$, indicating that the two datasets are consistent with being drawn from the same sample. The BAM study thus finds no correlation between the variables and the colour of a brown dwarf within each spectral type.

%__________________________________________________________________

\subsection{Binarity and variability}
The BAM sample includes 12 confirmed binaries out of 47 targets studied for binarity with another four SpeX spectra binary candidates. Including the binary candidates, 10 out of the 16 binaries in the BAM sample fall in the L/T transition. This is consistent with previous detections of an increase in the binary frequency across the L-T transition \cite{burgasser06}. Amongst the BAM variables, only 2M2255 is a confirmed binary, while 2M1207 and 2M2139 are binary candidates. Five of the variables are confirmed to be single, and six have not been studied for binarity. The limited data provides no evidence to support a correlation between variability and binarity amongst the objects in the BAM survey.

%__________________________________________________________________

\subsection{Persistence of variability}
\label{Persistence}

A recent multi-epoch ($\sim$4~years) monitoring study of the variable brown dwarf SIMP0136 \citep{metchev13} revealed that the target has significant evolution in its light curve, changing from highly variable to constant in a 2~month period. When compared to the SofI light curve for SIMP0136, the target shows a fascinating variation in amplitude. It appears to be variable at 3\% in the SofI data, while a month later it shows large amplitude variations ($\sim$9\%) in the $J$-band, only to appear constant a few months later. Similar night-to-night variations have also been seen in SDSS J105213.51+442255.7 \citep{girardin13}. The evolution indicates a lack of persistence in the source of variability over timescales longer than a few weeks and it suggests that the brown dwarfs identified as constant in this study might similarly exhibit periods of quiescence and enhanced activity. The BAM survey only examines variability on the timescale of a single rotation period or less as compared to some surveys \citep[e.g.][]{gelino02,enoch03} that study the flux variations of brown dwarfs over longer timescales. 

The BAM data, in combination with previous results, can be used to address the question of persistence of variability. Table~\ref{tabl:PersRes} summarizes the observations related to persistence of variability, using information presented in Table~\ref{tabl:variable_sample} and \ref{tabl:LitConst}. For greatest consistency with the BAM study, we consider other epochs of near-IR data rather than optical. A total of of 34 BAM targets have an earlier epoch of observation. 2M0228 is the only source measured to be variable in the optical ($I_c$) that switched from variable to constant. Table~\ref{tabl:PersRes} indicates that brown dwarf variability does not necessarily persist on longer timescales, with only half the BAM variables showing variation in both epochs. The survey finds four previously constant objects to be variable and nine targets previously reported as variable in the literature to be constant, making these ideal candidates for multiple epoch monitoring programs.

%__________________________________________________________________

\begin{table}
\centering
\caption{Summary of Persistence Results}
\label{tabl:PersRes}
\begin{tabular}{lc}
\hline
  Total targets with 2 epochs           &  34 \\
\hline
  Variable at 2 epochs                  &  6  \\
  Constant at 2 epochs                  &  15 \\
  Switch between variable and constant  &  13 \\
\hline

\end{tabular}
\end{table}
%__________________________________________________________________

%__________________________________________________________________

\section{Summary}
\label{conclusion}

We present the results of the largest near-IR brown dwarf variability survey conducted in the $J_s$-band using the NTT 3.5~m telescope. The BAM survey has an unbiased sample of 69 early-L through late-T brown dwarfs. A total of 14 variable objects were detected: six new variables not previously studied for variability, three objects previously reported as constant, and five previously known variables. The nine newly identified variables constitute a significant increase in the total number of known brown dwarf variables characterisable using ground-based facilities. In a recent study of 57 L4-T9 brown dwarfs \citep{radigan14}, a set of 35 targets were observed by both studies, enabling a direct comparison of results. Of the 35 targets in both samples, both studies classify a common 26 targets as not variable and a common four targets as variable. Of the five remaining variables noted in a single study (two in BAM, three in \citealt{radigan14}), four can be explained by differences in sensitivity for the specific light curves.

Rather than quoting a single number for the variability frequency, we discuss how the frequency of variable brown dwarfs depends on different factors such as the observed wavelength and the variability amplitude. The BAM study, representing the largest and most uniform ground-based search for variability, was designed to address the important question of the physical properties of brown dwarf atmospheres including the L-T transition. One class of models has suggested that this colour change, that defines the transition, may be a manifestation of the breakup of clouds resulting in a patchy coverage across the surface \citep{ackerman01}, which would have the observable consequence of enhanced variability at the L-T transition. Considering the results of this study, covering both transition and non-transition objects and statistical significance of the variability, there is no distinction between the variability frequency between the brown dwarfs in the transition region or outside the transition region. This suggests that the patchy cloud scenario may not provide the full explanation for the L-T transition or that the induced level of variability is substantially below the detection thresholds of the current study. The 14 variables, including the nine newly identified variables, will provide valuable systems with which to pursue additional questions of the physics of brown dwarf atmospheres, including the longitudinal and vertical variations of clouds and active regions which can be inferred from multi-wavelength follow-up monitoring.

%__________________________________________________________________

\begin{acknowledgements}
Based on observations made with ESO Telescopes at the La Silla Paranal Observatory under programme ID 188.C-0493. We would like to thank the anonymous referee for valuable suggestions thats helped improve this paper. PAW acknowledges support from STFC. JP was supported by a Leverhulme research project grant (F/00144/BJ), and funding from an STFC standard grant. This research has made use of the SIMBAD database and VizieR catalogue access tool, CDS, Strasbourg, France. The original description of the VizieR service was published in A\&AS 143, 23. This research has benefitted from the M, L, T, and Y dwarf compendium housed at DwarfArchives.org. This research made use of Astropy, a community-developed core Python package for Astronomy \citep{astropy}. We thank F. Pont, R. De Rosa and D. K. Sing for valuable feedback and discussion.
\end{acknowledgements}

%-------------------------------------------------------------------
\bibliography{myrefs}

\end{document}